\documentclass[epj,nopacs]{svjour}
\usepackage{xspace}
\usepackage{graphics}
\usepackage{graphicx}
\usepackage{amsmath}
\usepackage{booktabs}
\usepackage{xcolor}
\usepackage{colortbl} 
\usepackage{comment}
\usepackage{microtype}
\usepackage[normalem]{ulem}
\usepackage{newtxtext,newtxmath}
\usepackage[numbers,sort&compress]{natbib}
\usepackage[colorlinks=true, allcolors=blue]{hyperref}
\usepackage{array}
\usepackage{makecell}
\newcolumntype{L}[1]{>{\raggedright\arraybackslash}p{#1}}
\newcommand{\rev}[1]{{\color{black}#1}}
\newcommand{\revv}[1]{{\color{black}#1}}
\newcommand{\Tr}{\ensuremath{T^{}_{r}}\xspace}
\newcommand{\muBohr}{\ensuremath{\mu^{}_\text{B}}\xspace}

\begin{document}

\newcommand{\QEDref}{Aoyama:2012wk,Volkov:2019phy,Volkov:2024yzc,Aoyama:2024aly,Parker:2018vye,Morel:2020dww,Fan:2022eto}

\newcommand{\EWref}{Czarnecki:2002nt,Gnendiger:2013pva,Ludtke:2024ase,Hoferichter:2025yih}

\newcommand{\HVPref}{RBC:2018dos,Giusti:2019xct,Borsanyi:2020mff,Lehner:2020crt,Wang:2022lkq,Aubin:2022hgm,Ce:2022kxy,ExtendedTwistedMass:2022jpw,RBC:2023pvn,Kuberski:2024bcj,Boccaletti:2024guq,Spiegel:2024dec,RBC:2024fic,Djukanovic:2024cmq,ExtendedTwistedMass:2024nyi,MILC:2024ryz,FermilabLatticeHPQCD:2024ppc,Keshavarzi:2019abf,DiLuzio:2024sps,Kurz:2014wya}

\newcommand{\HLbLref}{Colangelo:2015ama,Masjuan:2017tvw,Colangelo:2017fiz,Hoferichter:2018kwz,Eichmann:2019tjk,Bijnens:2019ghy,Leutgeb:2019gbz,Cappiello:2019hwh,Masjuan:2020jsf,Bijnens:2020xnl,Bijnens:2021jqo,Danilkin:2021icn,Stamen:2022uqh,Leutgeb:2022lqw,Hoferichter:2023tgp,Hoferichter:2024fsj,Estrada:2024cfy,Ludtke:2024ase,Deineka:2024mzt,Eichmann:2024glq,Bijnens:2024jgh,Hoferichter:2024bae,Holz:2024diw,Cappiello:2025fyf,Colangelo:2014qya,Blum:2019ugy,Chao:2021tvp,Chao:2022xzg,Blum:2023vlm,Fodor:2024jyn}

\newcommand{\SMref}{Aoyama:2012wk,Volkov:2019phy,Volkov:2024yzc,Aoyama:2024aly,Parker:2018vye,Morel:2020dww,Fan:2022eto,Czarnecki:2002nt,Gnendiger:2013pva,Ludtke:2024ase,Hoferichter:2025yih,RBC:2018dos,Giusti:2019xct,Borsanyi:2020mff,Lehner:2020crt,Wang:2022lkq,Aubin:2022hgm,Ce:2022kxy,ExtendedTwistedMass:2022jpw,RBC:2023pvn,Kuberski:2024bcj,Boccaletti:2024guq,Spiegel:2024dec,RBC:2024fic,Djukanovic:2024cmq,ExtendedTwistedMass:2024nyi,MILC:2024ryz,FermilabLatticeHPQCD:2024ppc,Keshavarzi:2019abf,DiLuzio:2024sps,Kurz:2014wya,Colangelo:2015ama,Masjuan:2017tvw,Colangelo:2017fiz,Hoferichter:2018kwz,Eichmann:2019tjk,Bijnens:2019ghy,Leutgeb:2019gbz,Cappiello:2019hwh,Masjuan:2020jsf,Bijnens:2020xnl,Bijnens:2021jqo,Danilkin:2021icn,Stamen:2022uqh,Leutgeb:2022lqw,Hoferichter:2023tgp,Hoferichter:2024fsj,Estrada:2024cfy,Deineka:2024mzt,Eichmann:2024glq,Bijnens:2024jgh,Hoferichter:2024bae,Holz:2024diw,Cappiello:2025fyf,Colangelo:2014qya,Blum:2019ugy,Chao:2021tvp,Chao:2022xzg,Blum:2023vlm,Fodor:2024jyn}

\newcommand{\expref}{Muong-2:2025xyk,Muong-2:2023cdq,Muong-2:2024hpx,Muong-2:2021ojo,Muong-2:2021vma,Muong-2:2021ovs,Muong-2:2021xzz,Muong-2:2026qnz}

\title{CANTON-$\mu$ Proposal: A Next-Generation Muon $g\!-\!2$ Measurement at Sub-0.1~ppm Precision}
%\author{First author\inst{1} \and Second author\inst{2}% etc
%}                     % Do not remove
%

\author{
Ce Zhang\inst{1,}\thanks{\textcolor{black}{email: \texttt{ce.zhang@liverpool.ac.uk}}} \and
Yu Xu\inst{2,3,}\thanks{\textcolor{black}{email: \texttt{xuyu@gdlhz.ac.cn}}} \and
On Kim\inst{4,}\thanks{\textcolor{black}{email: \texttt{onkim@uw.edu}}} \and
Bingzhi Li\inst{5} \and
Zejia Lu\inst{7} \and
Saskia Charity\inst{1} \and
Guodong Shen\inst{2,3,6} \and
Liangwen Chen\inst{2,3,6} \and
Fedor Ignatov\inst{1} \and
Liang Li\inst{7} \and
Qiang Li\inst{8} \and
Xueheng Zhang\inst{2,3,6} \and
Zhiyu Sun\inst{2,3,6}
}

%\email{onkim@uw.edu}

\institute{
University of Liverpool, Liverpool, United Kingdom
\and
Advanced Energy Science and Technology Guangdong Laboratory, Huizhou 516000, China \and
Institute of Modern Physics, CAS, Lanzhou 730000, China \and
University of Washington, Seattle, Washington, USA \and
Zhejiang Lab, Hangzhou, China \and
School of Nuclear Science and Technology, University of Chinese Academy of Sciences, Beijing 100049, China \and
School of Physics and Astronomy, Shanghai Jiao Tong University, Shanghai, China \and
State Key Laboratory of Nuclear Physics and Technology, School of Physics, Peking University, Beijing 100871, China
}
%\institute{Insert the first address here \and the second here}
%
\date{Received: date / Revised version: date}
\abstract{
We propose a next-generation precision measurement of the muon anomalous magnetic moment (muon $g\!-\!2$) at the High Intensity Heavy-Ion Accelerator Facility (HIAF) in Huizhou, China.
We refer to this proposed experimental programme as \textbf{CANTON-$\mu$} (\emph{Coherent Anomalous magNetic momenT ObservatioN with muon}).
\rev{HIAF's intense, pulsed GeV-scale muon beams, particularly for negative muons, provide a promising basis for this programme.
Building on two previously proposed storage-ring concepts, this work develops HIAF-specific experimental schemes that relax the conventional magic-momentum constraint and allow greater flexibility in the choice of beam momentum.
We assess the expected muon intensity at HIAF and the corresponding statistical sensitivity.
For each scheme, we identify its distinctive systematic effects, examine feasible control strategies, and propose quantitative systematic-uncertainty targets.
Together, the statistical projections and systematic-uncertainty targets indicate prospective total precisions of 0.1~ppm in Phase-I, comparable to the current Fermilab precision, with the potential to reach the $0.05~\mathrm{ppm}$ level in Phase-II following the planned HIAF upgrade.}
At the ultimate projected precision, the measurement would provide a stringent test of the Standard Model and probe new physics at multi-TeV scales.
A focus on negative-muon measurements would enable direct comparison with existing high-precision positive-muon results and strengthen tests of CPT symmetry in the muon sector within the Standard-Model Extension, with a projected sensitivity at the $10^{-24}\ \mathrm{GeV}$ level, an order of magnitude beyond current limits.
}

%describes novel experimental approaches based on HIAF’s intense pulsed muon beams at the GeV scale, particularly for negative-muon polarity. 
%This work presents a detailed study of the expected muon beam properties at HIAF, establishing the statistical reach and level of systematic control required to achieve a precision of 0.1~ppm in Phase~1, matching the current Fermilab precision, and 0.05~ppm in Phase~2 with the HIAF upgrade.
%
%\keywords{Muon $g\!-\!2$, HIAF, Proton co-magnetometer, Weak-focusing storage ring}

\maketitle

\section{Introduction}
~\label{sec:intro}
The anomalous magnetic moment of the muon, commonly known as the muon `$g\!-\!2$’, is defined as $a_\mu \equiv (g_\mu-2)/2$, where $g_\mu$ denotes the muon $g$ factor. It is one of the most precisely predicted and experimentally measured quantities in particle physics. 
\rev{As a loop-induced quantum observable, muon $g\!-\!2$ receives contributions from all sectors of the Standard Model (SM) and probes physics across a wide range of energy scales.}
Since Schwinger’s pioneering calculation of its leading-order QED contribution~\cite{Schwinger:1951nm}, increasingly precise theoretical predictions and experimental measurements have established the muon $g\!-\!2$ as a ``golden observable'' for testing the internal consistency of the SM and probing physics beyond it.

Over the past decades, successive experiments at CERN~\cite{CERN-Mainz-Daresbury:1978ccd}, Brookhaven~\cite{Muong-2:2006rrc}, and most recently Fermilab~\cite{\expref}, have progressively refined the measurement of muon $g\!-\!2$. These efforts culminated in the 2025 Fermilab result~\cite{Muong-2:2025xyk} which reached a precision of 0.127~ppm and showed excellent agreement with the earlier results, providing a consistent experimental foundation for comparison with theory.
On the other hand, the SM prediction for the muon $g\!-\!2$ remains unsettled. Two comprehensive reviews released in 2020~\cite{Aoyama:2020ynm} and 2025~\cite{Aliberti:2025beg} by the Muon $g\!-\!2$ Theory Initiative, commonly referred to as the WP20 and WP25 white papers, have reported different numerical results for the SM prediction. The 2025 result is compiled based on Refs.~\cite{\SMref}. However, owing to the methodological differences between the 2020 and 2025 analyses, the 2020 result is not obsolete and should still be included in the discussion.

In general, the SM prediction consists of QED, electroweak, and hadronic contributions. The QED~\cite{\QEDref} and electroweak~\cite{\EWref} terms are known with sub-5 ppb precision, while the overall theoretical uncertainty is dominated entirely by the hadronic corrections, namely the hadronic vacuum polarization (HVP)~\cite{\HVPref} and hadronic light-by-light (HLbL)~\cite{\HLbLref} contributions governed by the strong interaction. These quantities involve low-energy, multi-scale QCD dynamics that cannot be treated perturbatively.
%are intrinsically difficult to compute from first principles, as they
Two complementary approaches are being pursued to calculate the HVP and HLbL contributions: data-driven (dispersive) methods~\cite{Colangelo:2014dfa,Colangelo:2014pva,Colangelo:2015ama} and lattice QCD~\cite{Blum:2014oka,Blum:2017cer,Asmussen:2022oql}. The data-driven approach determines the HVP through dispersion integrals of experimentally measured \(e^+e^-\to\mathrm{hadrons}\) cross sections, supplemented at higher energies by perturbative QCD input.
\revv{Comprehensive data-driven evaluations have been performed by several
groups~\cite{Davier:2019can,Davier:2023fpl,Keshavarzi:2018mgv,Keshavarzi:2019abf,Keshavarzi:2024wow,Colangelo:2018mtw,Leplumey:2025kvv,Colangelo:2022prz}.}
\rev{Despite its remarkable precision, unresolved tensions remain among the input datasets, including those between BaBar and KLOE~\cite{Aoyama:2020ynm,BaBar:2009wpw,BaBar:2012bdw,BESIII:2015equ,Aulchenko:2006dxz,CMD-2:2006gxt,CMD-2:2005mvb,CMD-2:2003gqi,CMD-2:2001ski,SND:2020nwa} and the more recent CMD-3 measurement~\cite{CMD-3:2023rfe,CMD-3:2023alj}. 
\revv{Hybrid approaches combining lattice QCD and experimental data have also made progress towards more precise HVP determinations~\cite{Boccaletti:2024guq}.}
These discrepancies directly constrain the precision achievable within the dispersive framework.} Lattice QCD, by contrast, provides an \emph{ab initio} framework to evaluate, in particular, the leading-order hadronic vacuum polarization (LO-HVP) directly from the QCD Lagrangian. Owing to substantial progress in numerical precision and in the control of systematic uncertainties, recent lattice-QCD results have reached a level of accuracy comparable to that of the data-driven determinations.

%The first is the data-driven (dispersive) method~\cite{Colangelo:2014dfa,Colangelo:2014pva,Colangelo:2015ama}, which determines the HVP through dispersion integrals of experimentally measured \(e^+e^-\to\mathrm{hadrons}\) cross sections, supplemented at higher energies by perturbative QCD input. This approach has achieved remarkable precision; however, limitations in internal consistency have emerged, initially between the BaBar and KLOE datasets~\cite{Aoyama:2020ynm}. More recently, the CMD-3 measurement of the \(e^+e^-\to\pi^+\pi^-\) channel in 2024~\cite{CMD-3:2023rfe,CMD-3:2023alj} yields a result that differs significantly from all previous \(e^+e^-\) measurements, introducing an additional tension whose origin remains unclear. These persistent discrepancies directly constrain the ultimate precision achievable within the dispersive framework.

%The second approach is lattice QCD~\cite{Blum:2014oka,Blum:2017cer,Asmussen:2022oql}, which provides an \emph{ab initio} framework to evaluate, in particular, the leading-order hadronic vacuum polarization (LO-HVP) directly from the QCD Lagrangian. Owing to substantial progress in numerical precision and in the control of systematic uncertainties, recent lattice-QCD results have reached a level of accuracy comparable to that of the data-driven determinations.

\begin{figure}[ht]
\centering
\includegraphics[width=0.99\linewidth]{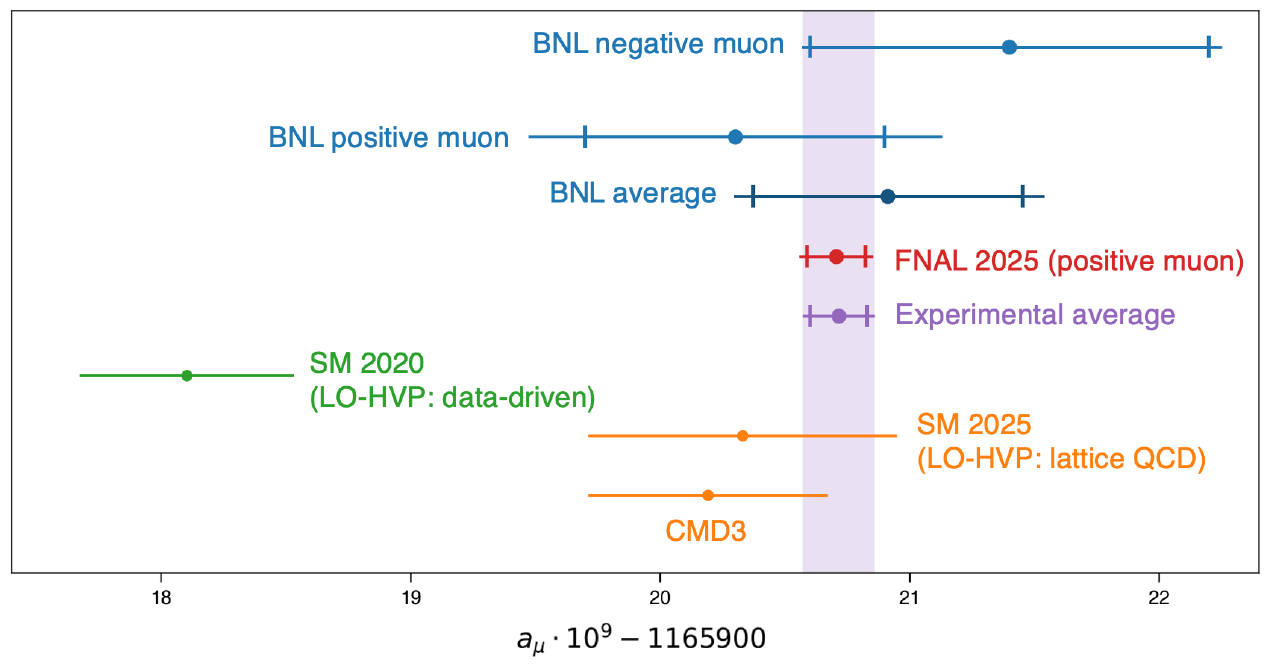}
\caption{\label{fig:gm2}Status of muon $g\!-\!2$ experimental measurements and SM predictions, with the latter showing results from both data-driven (green) and lattice QCD (orange) calculations of the LO-HVP. \revv{The CMD-3-based point highlights the present tension among the dispersive inputs.} }
\end{figure}

In light of the tensions affecting the dispersive method, the latest recommendation in 2025 adopts the lattice-QCD average for the leading-order HVP as the basis of the SM prediction. In both the 2020 and 2025 white papers, higher-order HVP contributions are taken from data-driven evaluations, while the hadronic light-by-light term is obtained from an average of data-driven and lattice-QCD results. 
%Currently, beyond the internal inconsistencies within the dispersive datasets, the most critical issue is the difference between the two theoretical frameworks themselves. 
Whereas the earlier 2020 white paper implied a noticeable deviation from experiment, the latest WP2025 finds no significant discrepancy with the experimental world average. \rev{These experimental and theoretical results are summarized in Fig.~\ref{fig:gm2}.}

In the coming years, ongoing theoretical developments are expected to provide critical clarifications and further reduce the overall uncertainty. Recent comprehensive reviews (e.g. \cite{Hertzog:2025ssc}) suggest that the theoretical precision could realistically approach the current experimental sensitivity, at the level of $0.13\ \mathrm{ppm}$. The MUonE collaboration at CERN~\cite{MUonE:2019qlm} also offers an independent method to calculate the hadronic vacuum polarization (leading-order) contribution, $a_\mu^{\rm HVP,\ LO}$. Once theory reaches comparable or even sub-0.1-ppm precision, it becomes essential to establish an independent measurement of the muon $g\!-\!2$ to enable a robust assessment of any observed deviations and to guide the future exploration of this benchmark quantity.

Experimentally, 
all muon $g\!-\!2$ experiments since CERN-III~\cite{CERN-Mainz-Daresbury:1978ccd,CERNMuonStorageRing:1977bbe} have relied on essentially the same methodology: injecting muons at the so-called “magic momentum” of about 3.1~GeV/$c$ into a continuous magnetic storage region, using electric quadrupoles (EQs) for vertical focusing, and calibrating the magnetic field with fixed and movable nuclear magnetic resonance (NMR) probes. 
In a storage region with magnetic field $\vec B$ and electric field $\vec E$, 
the difference between the \rev{muon} spin-precession and cyclotron frequencies is  
\begin{equation}
\begin{split}
\vec{\omega}_a
= -\,\frac{q}{m_\mu}\!\Biggl[
a_\mu\,\vec B
-a_\mu(\frac{\gamma}{\gamma+1})(\vec\beta\cdot\vec B)\vec\beta \\
-\left(a_\mu-\frac{1}{\gamma^2-1}\right)\frac{\vec\beta\times\vec E}{c}
\Biggr].
\end{split}
\label{eq:wa_general}
\end{equation}

Here $\gamma$ is the Lorentz factor and $\vec\beta=\vec v/c$. 
\rev{For muons moving in the horizontal plane in a vertical magnetic field, $\vec\beta\cdot\vec B=0$.}
The terms related to the
electric dipole moment (EDM) are neglected
in the equation.
Choosing the magic Lorentz factor, $\gamma_{\rm mag}\approx29.3$, corresponding to the magic momentum $p_{\rm mag}\approx3.094~\mathrm{GeV}/c$ suppresses the impact of the electric-field term, bringing it close to zero,
yielding the simplified form
\begin{equation}
\vec{\omega}_a \;\simeq\; -\,\frac{q}{m_\mu}\,a_\mu\,\vec B.
%\qquad (\text{magic }p).
\label{eq:wa_pureB}
\end{equation}
\rev{This is conventionally referred to as the magic-$\gamma$ condition.}

Operationally, the experiment determines $\omega_a$ from the time spectrum of high-energy decay electrons/positrons 
(via the parity-violating decay asymmetry)~\cite{Muong-2:2021vma}.
\rev{The extracted frequency is subject to beam-dynamics corrections~\cite{Muong-2:2021xzz}, including those arising from the residual electric-field ($\vec{\beta}\times\vec{E}$ term) and vertical-pitch ($\vec{\beta}\cdot\vec{B}\neq0$) effects associated with departures from the ideal beam configuration.}
%\rev{with necessary beam-dynamics corrections~\cite{Muong-2:2021xzz} such as due to the residual electric-field ($\vec\beta\times\vec E$ term) and vertical pitch (\rev{$\vec\beta\cdot\vec B\neq0$}) effects away from the ideal beam configuration; 
%these are sub-ppm but enter the final error budget.}

\rev{
The magnetic field is measured using proton nuclear magnetic resonance (NMR)~\cite{Muong-2:2021ovs}. A mobile trolley equipped with NMR probes periodically maps the field throughout the SR, while fixed probes track its temporal evolution between trolley measurements. Following absolute calibration and weighting by the muon distribution, these measurements determine the muon-weighted proton precession frequency $\widetilde{\omega}_p^\prime(T_r)$, which is related to the effective magnetic field experienced by the muons through
\begin{equation}
\widetilde{B}
=
\frac{\hbar\,\widetilde{\omega}_p^\prime(T_r)}
     {2\mu_p^\prime(T_r)},
\label{eq:nmr_field}
\end{equation}
where $\widetilde{B}=\langle B\times M\rangle$ denotes the magnetic field averaged over the reconstructed muon distribution $M$, and $\mu_p^\prime(T_r)$ is the magnetic moment of a shielded proton in the reference water sample at temperature $T_r$.

Combining the experimentally determined frequency ratio with the definition of the Bohr magneton, \revv{$\muBohr=e\hbar/(2m_e)$}, the muon anomalous magnetic moment can be expressed as
\begin{equation}
a_\mu
=
\frac{\omega_{a}}
     {\widetilde{\omega}_p^\prime(\Tr)}
\frac{\mu'_\text{p}(\Tr)}{\muBohr} \frac{m_{\mu}}{m_\text{e}},
\label{eq:a_mu_extraction_fnal}
\end{equation}
where the ratios $\mu'_\text{p}(\Tr)/\muBohr$ and $m_{\mu}/m_e$ are \revv{external inputs. Their product is precisely known to 22~ppb from other experiments, dominated by the uncertainty in $m_\mu/m_e$}~\cite{mohr2024codatarecommendedvaluesfundamental}.
}
%$\mu_e(H)$ and $\mu_e$ are the magnetic moments of an electron bound in hydrogen and a free electron, respectively, and $g_e$ is the electron $g$ factor. The frequency ratio is determined by the experiment, while the remaining factors are obtained from external precision measurements.

%One then forms the frequency ratio $R=\omega_a/\omega_p$ and uses the known magnetic-moment ratio $\lambda=\mu_\mu/\mu_p$ to extract $a_\mu$:
%\begin{equation}
%a_\mu = \frac{R}{\lambda - R},
%\qquad R \equiv \frac{\omega_a}{\omega_p}.
%\label{eq:a_mu_extraction}
%\end{equation}

\rev{While this conventional approach has been extraordinarily successful, its reliance on the magic-$\gamma$ condition fixes the muon momentum, limiting the flexibility to operate at higher muon energies.
This limitation is particularly relevant to the statistical precision of $\omega_a$}, which can be expressed as~\cite{Muong-2:2007tpa}
\begin{equation}
\frac{\Delta\omega_a}{\omega_a}
=
\frac{1}{\omega_a\gamma\tau P}
\sqrt{\frac{2}{NA^2}} ,
\label{eq:sensitivity}
\end{equation}
where $\tau$ is the muon lifetime at rest, $P$ is the beam polarization, $N$ is the number of analyzed decay electrons or positrons, and $A$ is the effective decay asymmetry. Since $\omega_a \propto B$ (Eq.~\eqref{eq:wa_pureB}), the statistical sensitivity scales linearly with $B$, $\gamma\tau$, and with the square root of the number of events $\sqrt{N}$. \rev{This relation highlights that, beyond increasing the number of detectable muons, raising the muon energy—and hence $\gamma$—and operating at a higher magnetic field provides an efficient route to improving statistical precision. The polarization $P$ is generally expected to be high for muon beams selected from forward pion decays: at Fermilab, for example, a polarization of $P\simeq95\%$ was achieved~\cite{Muong-2:2021xzz}. The effective asymmetry $A$ is optimized primarily at the downstream-analysis stage through positron-energy selection or weighting. Therefore neither $P$ nor $A$ is treated as a primary optimization variable in the present discussion of the upstream beam and storage-ring configuration.}

\rev{Operation at higher muon energies requires abandoning the conventional magic-$\gamma$ condition. A new experiment must therefore suppress the electric-field correction arising from the $\vec{\beta}\times\vec{E}$ term in Eq.~\eqref{eq:wa_general}, which constitutes one of the dominant sources of systematic uncertainty. One possible approach is to completely eliminate electric fields from the muon storage region. Doing so, however, also removes the electric focusing conventionally used to confine the stored muon beam, including the vertical focusing provided by the EQs at Fermilab. This challenge is particularly important because muons produced as secondary particles from pion decay typically have a large emittance. A viable experiment must therefore employ either a novel focusing scheme that does not rely on electric fields, or a high-quality muon beam that can be stored without conventional electric focusing.} The latter approach is being pursued by the J-PARC Muon $g\!-\!2$/EDM collaboration~\cite{abe_new_2019}, which is developing a technique to produce and re-accelerate ultracold positive muons with low emittance.
Although innovative, this approach currently has a relatively low overall muon efficiency. Its Phase-I design precision of approximately $0.45$~ppm for $a_\mu$ remains less precise than the Fermilab benchmark, \rev{and studies toward higher sensitivity are ongoing~\cite{Venanzoni:2025rsd}}.

It is also noteworthy that the Fermilab experiment measured $g\!-\!2$ of only the positive muon ($\mu^+$), whereas the earlier Brookhaven E821 experiment measured both charge states, as shown in Fig.~\ref{fig:gm2}. The present global sensitivity is overwhelmingly driven by the Fermilab $\mu^+$ data.
The E821 results for $\mu^+$ and $\mu^-$ were found to be consistent within their respective uncertainties, providing an important test of Charge–Parity–Time (CPT) symmetry~\cite{Bluhm:1999dx,Muong-2:2007ofc, Gomes:2014kaa}. 
Within the Standard-Model Extension (SME), comparisons between the two charge states can constrain coefficients associated with CPT and Lorentz violation~\cite{Bluhm:1999dx,Kostelecky:2008ts}. In the absence of a modern $\mu^-$ $g\!-\!2$ measurement, these constraints have not been updated over the past two decades.
Moreover, the J-PARC muon $g\!-\!2$ approach is intrinsically limited to positive muons because it relies on the laser ionization of muonium~\cite{Zhang:2022ilj}. A new muon $g\!-\!2$ programme capable of matching or surpassing the current Fermilab precision would be particularly valuable for the negative muon.

At present, among the very few facilities worldwide, China's High Intensity Heavy-Ion Accelerator Facility (HIAF)~\cite{Yang:2013yeb,Yang:2023hfx} stands out with a dedicated plan to produce GeV-scale pulsed, high-intensity muon source of both polarities using high-energy proton and ion beams. This unique capability makes HIAF a promising site for a next-generation muon precision measurement.

\rev{This work presents a site-specific assessment of the feasibility of a muon $g\!-\!2$ programme at HIAF. We first examine the expected muon beam properties and evaluate whether the achievable beam intensity can support a precision measurement (Sec.~\ref{sec:HIAF}). We then adapt and assess two previously proposed muon storage ring (SR) concepts for the HIAF beam conditions (Sec.~\ref{sec:design}) and evaluate their achievable sensitivities (Sec.~\ref{sec:sensitivity}). We identify a different balance of systematic effects in the resulting SR designs and assess control strategies and systematic-uncertainty targets consistent with prospective total precisions of 0.1~ppm in Phase-I and 0.05~ppm in Phase-II. Finally, we assess the physics reach enabled by the projected sensitivities (Sec.~\ref{sec:newphysics}), including the new-physics scales that could be probed and the sensitivity to CPT violation in the muon sector.}

%A new experiment is therefore crucial to measure the muon $g\!-\!2$ with a precision matching or surpassing the current Fermilab record, with particular emphasis on the negative muon ($\mu^-$). Approaches employing different focusing concepts or magnet geometries would naturally introduce distinct systematic effects, providing an essential cross-check and strengthening the overall robustness of the global $g\!-\!2$ program.

\section{Muon Beam at HIAF}
\label{sec:HIAF}
\subsection{Muon Production Simulation for $g\!-\!2$}
\label{subsec:HIAFa}

HIAF is an advanced accelerator complex currently under construction since 2018 in Huizhou, Guangdong, China~\cite{Zhou:2022pxl}. 
Upon completion, it will provide intense heavy-ion beams across a broad energy range to support a diverse program of nuclear, particle, and applied-physics experiments~\cite{Xiaohong:2018weu,Chen:2024wad}. Commissioning of the first ion beam began at the end of 2025.

\begin{figure}[ht]
\centering
\includegraphics[width=0.99\linewidth]{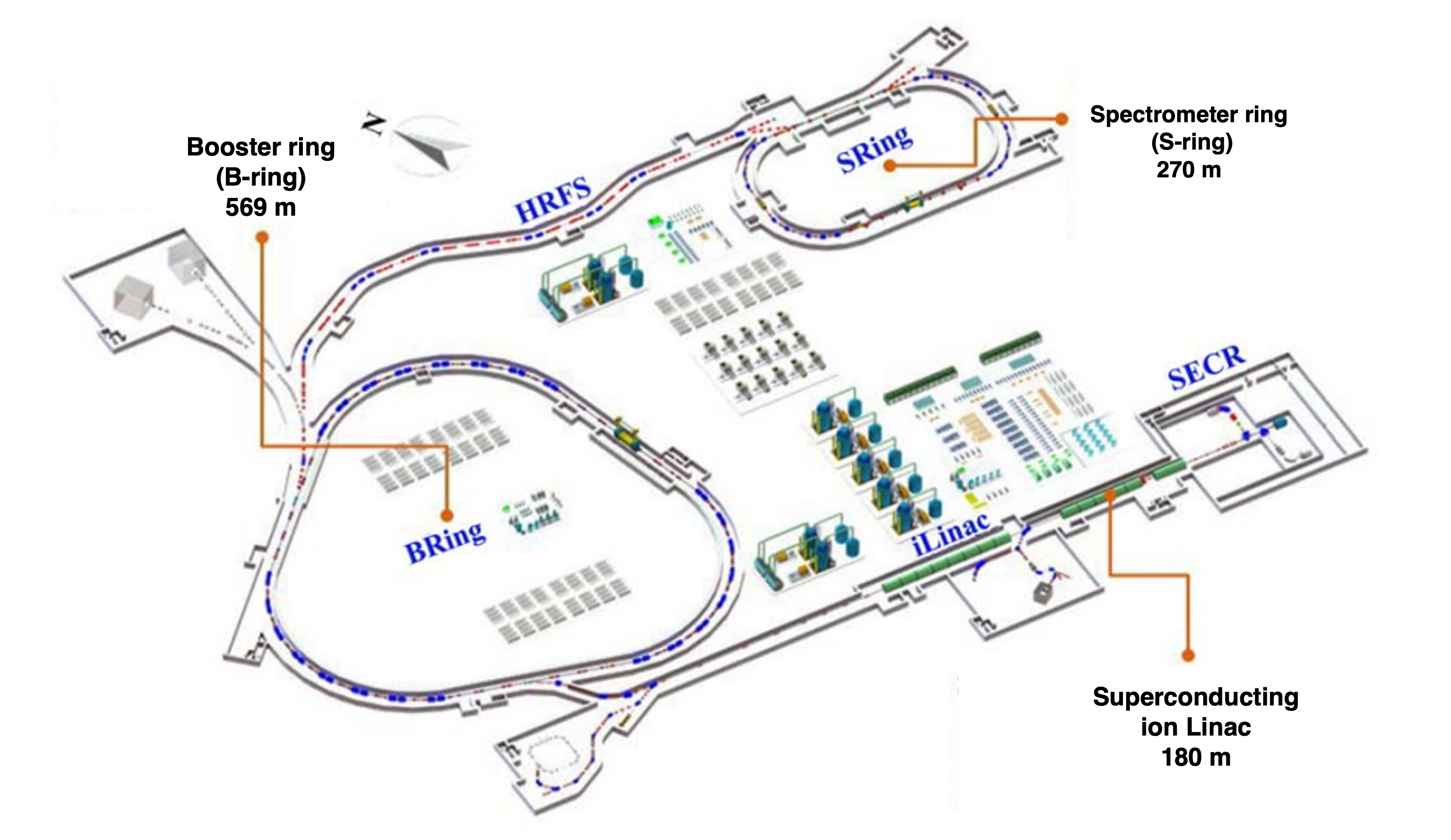}
\caption{\label{fig:HIAF_photo}The layout of HIAF, adapted from Fig.~1 in Ref.~\cite{Xiaohong:2018weu}.}
\end{figure}

The accelerator complex consists of a high-current ion source, a superconducting linear accelerator (iLinac), a 569~m booster synchrotron (BRing), a 192~m High-energy FRagment Separator (HFRS), and a high-precision storage ring (SRing), as illustrated in Fig.~\ref{fig:HIAF_photo}. The BRing can accelerate heavy ions up to hundreds to thousands of MeV/u, with fast- and slow-extraction modes enabling both pulsed and quasi-continuous operation, with intensities as large as $10^{11}$–$10^{12}$ particles per pulse (ppp)~\cite{Wang:2025sbe,Wang:2025scl}. The SRing is capable of storing beams with energies of several GeV per nucleon. For proton operation, the BRing will accelerate beams to approximately 9 GeV, 
enabling efficient production of secondary particles such as pions and muons from targets.

The HFRS beamline~\cite{Sheng:2023ojn,Wang:2025tpy} provides high-resolution separation and transport of secondary particles produced from high-energy projectile fragmentation and in-flight fission processes.
It comprises a multi-stage magnetic system, including a pre-separator and a main separator, with dipole and quadrupole groups configured for flexible operation modes optimized for either beam purification or spectroscopic studies.
With a maximum magnetic rigidity of 25~Tm, the HFRS can accommodate the full momentum range from pion decay at 9–25~GeV proton energies, enabling the transport of muons with momenta up to 7.5~GeV/$c$.
Its angular acceptance is $\pm30$~mrad (horizontal) and $\pm15$~mrad (vertical), with a longitudinal momentum acceptance of about $\pm2\%$.  
These parameters make the HFRS ideal not only for the separation of exotic radioactive nuclei but also exceptionally well suited for the production and transport of high-energy pions and muons.
The long flight path of about 190~m is comparable to the decay length of multi-GeV pions, enabling efficient pion-to-muon conversion within the beamline. Its large acceptance and tunable optics allow the momentum and phase-space properties of the muon beam to be optimized for different experimental requirements, such as high intensity or pulsed operation.

\begin{figure}[ht]
\centering
\includegraphics[width=0.9\linewidth]{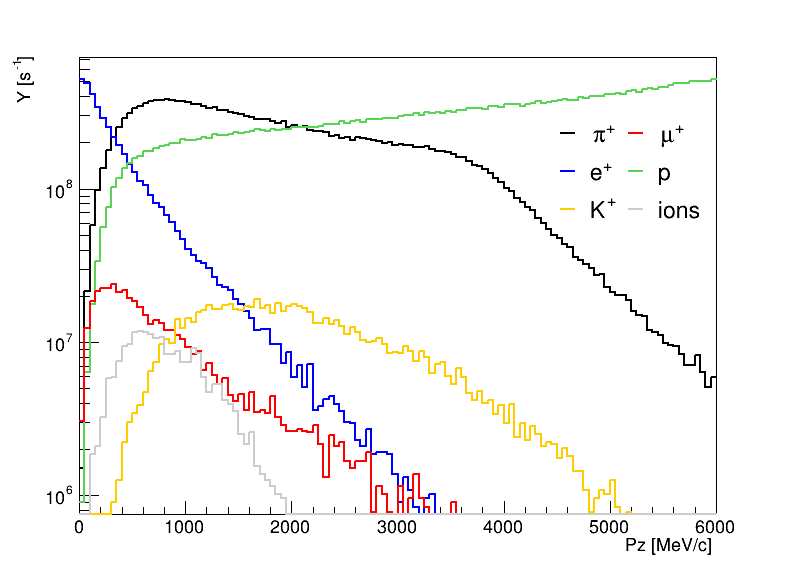}
\includegraphics[width=0.9\linewidth]{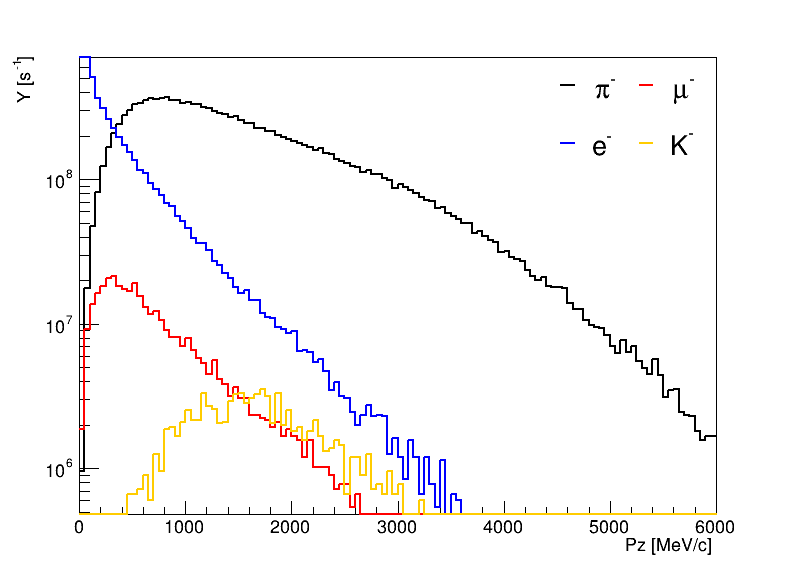}
\caption{\label{fig:pimuon} Simulated production rates of positive (top) and negative (bottom)
secondary particles as functions of the longitudinal momentum at
the HIAF production target for a $9.3~\mathrm{GeV}$ proton beam. The
distributions are shown before HFRS magnetic separation.}
\end{figure}

\begin{figure}[ht]
\centering
\includegraphics[width=0.99\linewidth]{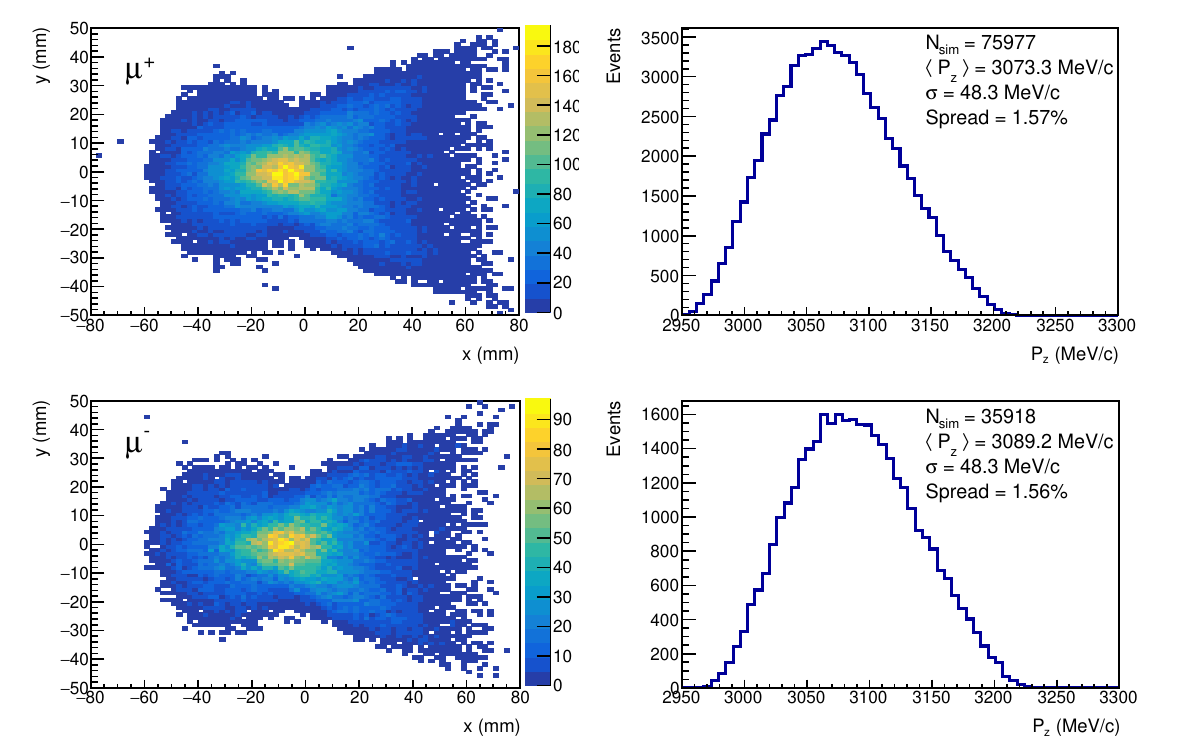}
\caption{\label{fig:profile} Simulated beam profiles (left) and momentum spread (right) for purified muons at 3.1~GeV/c, evaluated at the HFRS exit. \rev{Upper panels: positive muons; lower panels: negative muons.}}
\end{figure}

Since a precision measurement of the muon $g\!-\!2$ demands an exceptionally pure muon beam, we evaluate the performance of a pure muon beam, following the simulation of pion and muon production and purification strategy outlined in Ref.~\cite{xu_feasibility_2025}. The raw secondary beams contain background components from $\pi$, $p$, and $e$ contamination, as illustrated in Figure~\ref{fig:pimuon}. They are efficiently removed through the two-stage magnetic separation system of the HFRS. By tuning the rigidity difference $\Delta B\rho$ between the pre-separator and main separator to 0.13 Tm for $\mu^-$ and 0.17 Tm for $\mu^+$, \rev{the simulated residual hadronic background is at the level of one particle per $10^{5}$ muons for both polarities. For the representative positive and negative-muon beams selected at approximately $3~\mathrm{GeV}/c$, the simulated relative momentum spreads are approximately $1.57\%$ and $1.56\%$, respectively, while the transverse beam sizes range from $2$ to $10~\mathrm{cm}$ within 90\% containment, as shown in Fig.~\ref{fig:profile}.}

\rev{For the baseline $9.3~\mathrm{GeV}$ proton beam, simulations indicate that the most forward-going muons with momenta around $3~\mathrm{GeV}/c$ retain a longitudinal polarization of approximately $85\%$--$90\%$. The reduction arises mainly because some of the selected high-energy muons originate from higher-energy pion decays and do not correspond to the ideal forward-decay kinematic limit.} The time structure is inherited from the BRing extraction modes: either fast-pulsed operation at $3~\mathrm{Hz}$ or slow extraction with a two-second cycle and a one-second spill.

\begin{figure}[ht]
\centering
\includegraphics[width=0.9\linewidth]{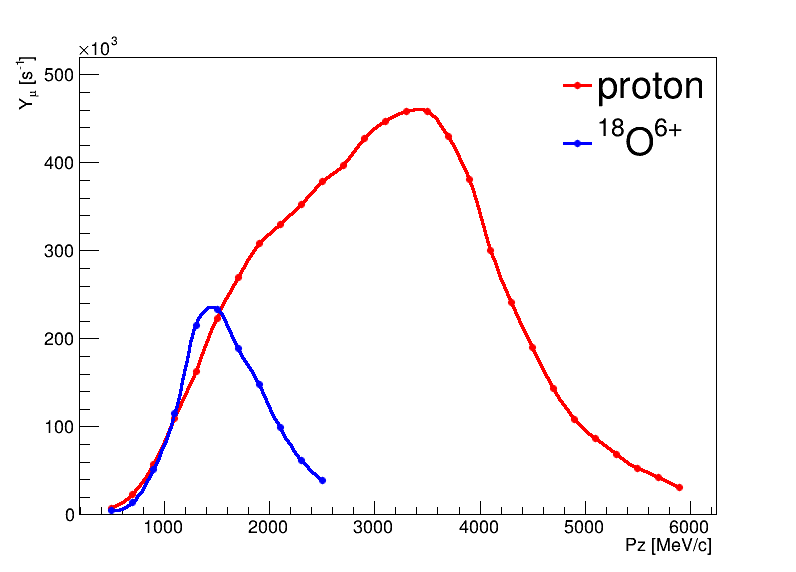}
\includegraphics[width=0.9\linewidth]{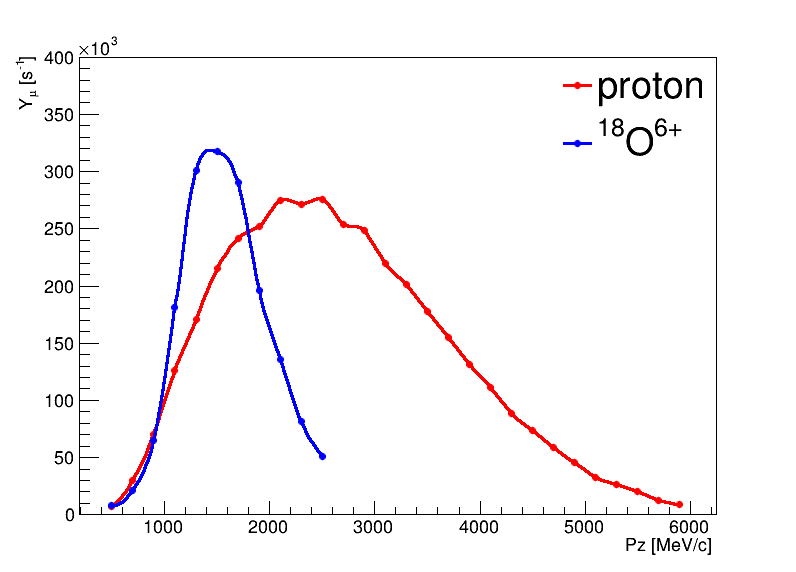}
\caption{\label{fig:intensity} Purified muon flux at the exit of the HFRS as a function of muon momentum. Each point represents the \rev{simulated} muon intensity for a specific momentum bin \rev{(top: positive muon; bottom: negative muon). The plotted rates retain the earlier beam-intensity normalization adopted in Ref.~\cite{xu_feasibility_2025}}. }
\end{figure}

Following the purification, the muon flux is substantially reduced but remains at a level suitable for the $g\!-\!2$ measurement. Figure~\ref{fig:intensity} shows the purified muon flux, in comparison with the unpurified flux from Fig.~5 of Ref.~\cite{xu_feasibility_2025}. 
The purified muon beam achieves output intensities of approximately $3.7\times10^{5}~\mu^-$/s (at 1.5~GeV/c corresponding to the \(\mathrm{^{18}O^{6+}}\)) and $4.5\times10^{5}~\mu^+$/s (at 3.5~GeV/c corresponding to the proton), respectively. At a beam momentum of roughly $3.1$~GeV/$c$, the negative-muon intensity is around $2.5\times10^{5}~\mu^-/\mathrm{s}$, compared with about $4.0\times10^{5}~\mu^+/\mathrm{s}$ for positive muons.

While the feasibility study in Ref.~\cite{xu_feasibility_2025} assumed a primary proton beam intensity of $2\times10^{12}$~ppp (\rev{$6\times10^{12}$ protons/s}) based on earlier design parameters, the HIAF beam intensity is now projected to reach approximately $5\times10^{13}$~s$^{-1}$ by 2026. For proton operation at 9.3 GeV, the calculated yields of \rev{purified} muons reach $3.8\times10^{6}$ $\mu^+$/s at 3.5 GeV/$c$ and $2.1\times10^{6}$ $\mu^-$/s at 2.3 GeV/$c$. Comparable intensities are obtained with light-ion projectiles such as $^{18}$O$^{6+}$ (2.6 GeV/u), which produce a few $10^{6}$ $\mu^-$/s in the 1–2 GeV momentum range. 

Remarkably, at HIAF the negative-muon beam reaches an intensity comparable to that of its positive counterpart, which is a notable contrast with conventional low-energy muon facilities, where the flux of surface muons ($p\approx28$~MeV/$c$), a primary source of $\mu^+$, is overwhelmingly higher than that of $\mu^-$ due to the intrinsic asymmetry of pion production and decay in the target.
This unique feature allows the $g\!-\!2$ experiment at HIAF to benefit from a high-quality negative-muon beam, offering additional physics potential as discussed in Section~\ref{sec:newphysics}.

Furthermore, a planned upgrade, HIAF-U, is foreseen for 2030 and beyond~\cite{ZHAO:2020llg}. In this upgrade, the HIAF heavy-ion superconducting linear injector will have its energy increased to 150–200 MeV/u, the BRing-N synchrotron will be transformed into a fast-cycling booster, and a fully superconducting main synchrotron, BRing-S, will be constructed in the same tunnel. The proton intensity is expected to reach $4\times10^{14}$/s, and the proton energy raised from 9 GeV to 25 GeV. The repetition rate of the BRing extraction is expected to increase from the current few-hertz operation to 10–12 Hz, further enhancing the average beam power delivered to the muon target. 
These upgrades would boost the muon flux by another order of magnitude, reaching approximately $10^{7}~\mu/\mathrm{s}$, while extending the accessible muon momentum to approximately $20~\mathrm{GeV}/c$.
\rev{Preliminary simulations further indicate that the most forward-going muons at momenta around $20~\mathrm{GeV}/c$ could retain a longitudinal polarization exceeding $90\%$.} Together, these upgrades would substantially expand HIAF's capabilities for precision muon physics, potentially establishing it as one of the world's leading high-energy muon facilities.
%This would position HIAF among the world's most powerful muon sources for precision physics, significantly extending its capabilities. 
Table~\ref{tab:beam-comparison} provides a quantitative comparison of the beam specifications at Fermilab (Ref.~\cite{Muong-2:2015tdr,Muong-2:2021xzz,Muong-2:2024hpx}), HIAF, and HIAF-U.

\begin{table*}[ht]
\centering
\caption{Comparison of muon beam parameters at Fermilab and HIAF.}
\begin{tabular}{p{5.5cm}p{2.8cm}p{2.8cm}p{2.8cm}}
\toprule
  & Fermilab & HIAF & HIAF-U\\
\midrule
Proton intensity ($\mathrm{s^{-1}}$) & $6.8\times10^{13}$ & $5\times10^{13}$ & $4\times10^{14}$ \\
Proton energy (GeV) & 8.0 & 9.1 & 25 \\
Repetition frequency (Hz) & \textcolor{black}{11--12} & 3 & 10--12 \\
Proton bunch time width (ns) & $\sim$\rev{120} & $\sim$100--400 & $\sim$100--400 \\
Muon flux after transport ($\mu$/s) & $5\times10^6~$ & $\sim4\times10^6~$ & $\sim10^7$--$10^8$ \\
Typical muon momentum (for $g\!-\!2$) & 3.1~GeV/$c$ & 2--4~GeV/$c$ & 10--20~GeV/$c$ \\
Muon momentum spread (\%) & 2 & 1.6 & Under study \\
\bottomrule
\end{tabular}
\label{tab:beam-comparison}
\end{table*}

\subsection{Implications of Beam Parameters for the Muon $g\!-\!2$ Design}
\label{subsec:HIAFb}

The optimal beam configuration for a precision muon $g\!-\!2$ measurement is a \emph{pulsed} beam with a large number of evenly spaced bunches, each having minimal time width. 
With proton intensity and energy comparable to Fermilab's, HIAF meets the essential prerequisite for a competitive muon flux.
The next most critical parameter is the \textbf{repetition frequency}. 
At a fixed total proton intensity per second, a higher repetition rate corresponds to a smaller proton population per pulse and therefore to a lower instantaneous muon flux per bunch and allows for more uniform data accumulation over time.
This is particularly beneficial for $g\!-\!2$, as excessive instantaneous rates can lead to severe pile-up effects in the positron detection system, requiring a careful detector segmentation and waveform reconstruction design. 
\rev{For Phase-I purified flux of approximately $(2$--$4)\times10^{6}$ muons/s and a repetition frequency of 3 Hz, the number of muons delivered toward the injection system is approximately $(0.7$--$1.3)\times10^{6}$ per fill. 
At Fermilab, the comparable flux of $5\times10^{6}$ muons/s at a repetition frequency of 11--12~Hz corresponded to approximately $4\times10^{5}$ muons per fill delivered toward the SR.}
For Phase-II, the planned repetition rate of 10--12~Hz for HIAF-U will be comparable to that at Fermilab.

The current \textbf{momentum spread} at HIAF is approximately 1.6\%, comparable to the roughly 2\% momentum spread of the muon beam delivered to the Fermilab muon $g\!-\!2$ storage region~\cite{Muong-2:2021xzz}.
At Fermilab, the stored beam was confined to a momentum spread of only about 0.2\% around the magic momentum~\cite{Froemming:2019gcn}. The narrow momentum acceptance of the injection and storage systems reduced the stored population to $\mathcal{O}(10^{4})$ muons per fill.
\rev{At Fermilab, this efficiency was also limited by the inflector--kicker injection system. Adopting a Fermilab-like momentum acceptance and injection efficiency therefore provides a conservative baseline for HIAF, corresponding to an expected stored population of approximately $10^{4}$--$10^{5}$ muons per fill. Therefore, the lower HIAF repetition rate of 3~Hz concentrates the events into fewer fills and increases the per-fill occupancy up to approximately three times that at Fermilab.
The resulting pile-up systematic uncertainty and detector requirements will be discussed in Sec.~\ref{sec:sensitivity}.}

Another key parameter is the \textbf{proton bunch time width}, which determines the temporal profile of injected muons in the SR. 
The time structure of the proton beam is largely inherited by the secondary muon beam,
with only a minor additional broadening introduced by the pion production and decay processes at the target. 
Therefore, the achievable proton bunch width directly defines the minimum time width of the muon bunches delivered to the experiment. 
%The conservative 400~ns bunch width assumed in Ref.~\cite{xu_feasibility_2025} can be reduced to 100~ns or even less, which is technically feasible at HIAF.
\rev{The $400~\mathrm{ns}$ bunch width assumed in Ref.~\cite{xu_feasibility_2025} provides a conservative baseline. Compression towards $100~\mathrm{ns}$ may be achievable at HIAF through dedicated longitudinal beam-dynamics and RF studies. We therefore regard $100~\mathrm{ns}$ as a prospective design target rather than a demonstrated capability.}
This flexibility is crucial because each bunch must be fully injected within one cyclotron period of the SR,
$T_c \propto \frac{\gamma m_\mu}{qB}$,
so that narrower bunches allow shorter $T_c$ and, consequently, operation at higher magnetic fields without overlap between injections, \rev{thereby improving the experimental precision.}

Given the muon momentum range of HIAF of 2--4~GeV/c, operation at a few-tesla magnetic field (beyond Fermilab’s 1.45~T) is realistic. 
This would correspond to a storage radius of a few meters (1--10~m), which remains practical from an engineering standpoint. 
At the upgraded HIAF-U, the higher muon energy (10--20~GeV/c) would lead to proportionally longer cyclotron periods and larger ring radii for the similar $B$ field. 
Such a configuration could be considered for a future phase, potentially utilizing the existing SRing as a muon storage ring~\cite{ZHAO:2020llg}, provided a cost-effective detector system can be realized. 
The higher beam energy at HIAF-U would enable an order-of-magnitude improvement in the attainable statistical precision.
%, as will be discussed further in Sec.~\ref{sec:sensitivity}.

\rev{Finally, we consider the residual hadronic contamination in the HIAF beam. The simulations predict a hadron-to-muon ratio of approximately $10^{-5}$ at the exit of the purification system. For the Phase-I muon flux and a repetition rate of 3~Hz, this corresponds to approximately 7--13 residual hadrons per fill before injection into the storage region. Assuming conservatively that the hadron-to-muon ratio is not enhanced during injection and storage, the upper-end population of $10^{5}$ stored muons per fill would imply only about one stored hadron per fill. This contamination level is comparable in order of magnitude to the Fermilab design beam, for which the residual pion fraction was below $10^{-5}$ and the proton component was largely removed in the Delivery Ring~\cite{Stratakis:2019sun}.}

%suggesting a similar injection efficiency of roughly 2\%. In the Fermilab experiment, this efficiency was largely limited by the stringent momentum acceptance requirement near the magic momentum~\cite{Muong-2:2021xzz}.
%At Fermilab, the final stored-beam momentum spread is only about 0.2\%, implying that a large fraction of upstream muons are rejected during injection~\cite{Froemming:2019gcn}. 
%Further investigation will explore whether part of the upstream phase space can be compressed toward the central momentum (e.g., 1.5–4~GeV/$c$ for negative muons) and whether a wider momentum acceptance could be accommodated in future designs, potentially improving the overall injection efficiency.

%Operating away from the muon’s magic momentum would unavoidably require novel experimental concepts, which we introduce below.

\section{Storage Ring Design Concepts}
\label{sec:design}

\rev{This section presents two storage-ring concepts proposed in earlier studies and discusses their possible application to a future muon $g\!-\!2$ experiment at HIAF. Both remain preliminary but offer possible routes to relaxing the constraints associated with conventional magic-momentum operation. The following subsections summarize the design principles, the key advantages and challenges of each concept. Their potential systematic uncertainties are considered in the subsequent section.}

\subsection{Concept A: Sector Magnet with Polarized Proton Co-Magnetometer}
\label{subsec:designA}

\begin{figure}[ht]
\centering
\includegraphics[width=0.98\linewidth]{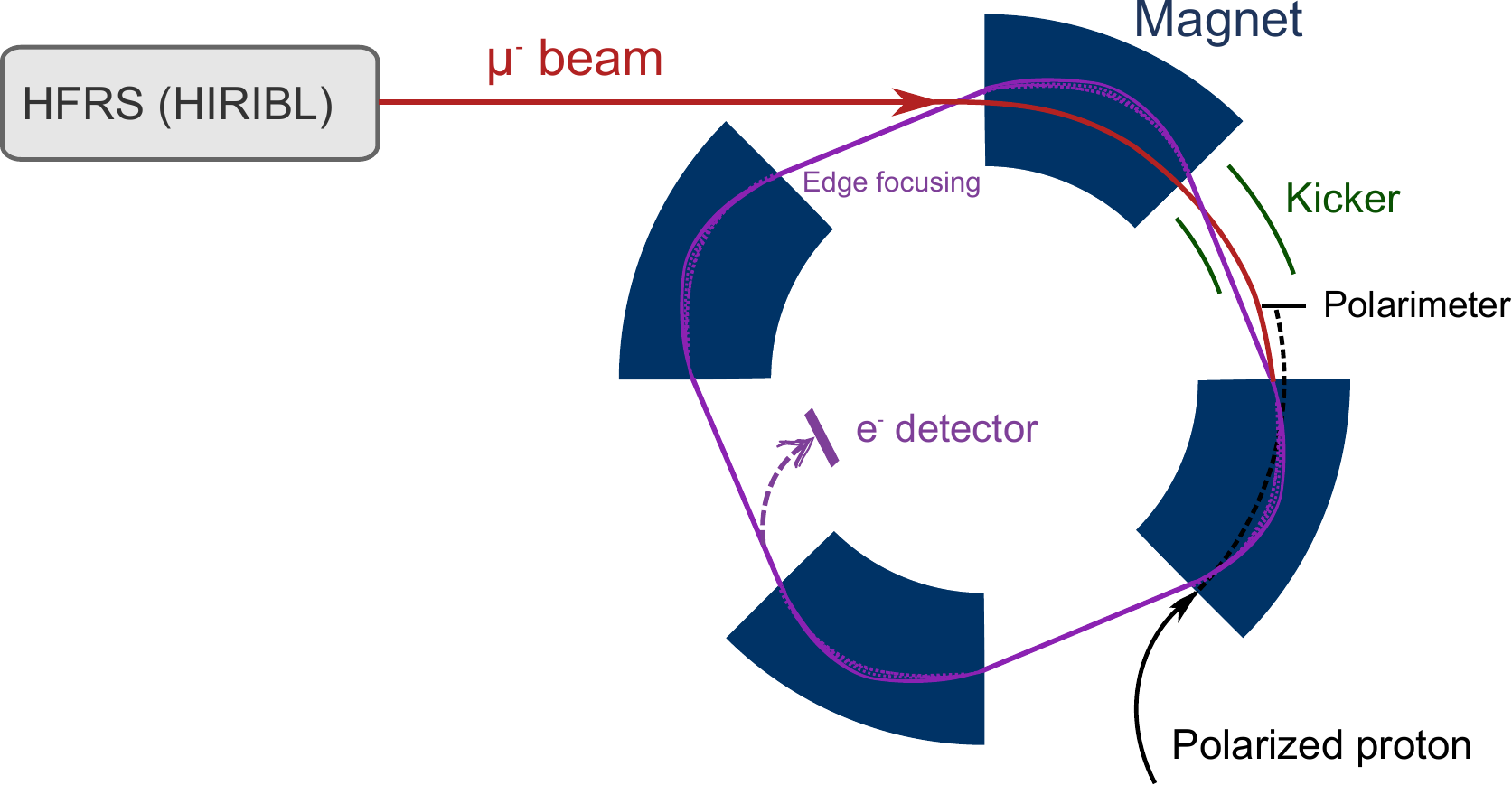}
\caption{\label{fig:designA}Schematic design of Concept~A with a polarized-proton co-magnetometer.}
\end{figure}

The discrete-sector muon storage region was first proposed by Farley in 2004~\cite{Farley:2003mj}. In this design, muons travel in straight lines through vacuum gaps between magnetic sectors and are bent only inside the dipoles, as illustrated in Fig.~\ref{fig:designA} for negative-muon injection (the configuration accommodates both polarities).
The natural edge fields of each sector provide vertical focusing (\emph{edge focusing}), eliminating the need for electrostatic quadrupoles.

In this configuration, the absence of electric fields removes the need for the conventional magic momentum. \rev{Measurements} can be performed over a wide range of \rev{muon} momenta accessible at HIAF, free from the electric-field-related systematic effects.
This flexibility in momentum choice is accompanied by a second key feature: while different muon momenta correspond to distinct orbital radii, the time-averaged magnetic field experienced by a muon over one full revolution is independent of its radius. This ensures that all muons sample the same average field, suppressing systematic uncertainties associated with precise orbit determination. 

In addition, the modular sector design naturally includes straight sections with essentially zero magnetic field. These field-free regions are ideal locations for injection systems and detectors for field calibration.
Muon injection can be performed using only kickers, avoiding the need for a dedicated inflector magnet, therefore simplifying the magnetic layout and eliminating the complex field-compensation scheme required in the Fermilab experiment.
\rev{Moreover, whereas the Fermilab kickers must provide a substantial deflection following passage through the inflector, the sector-ring geometry may allow the injection angle to be chosen such that, after traversing the entrance edge of the first sector magnet, the muon beam is already close to the desired closed orbit. The kicker would then provide a relatively small corrective deflection, potentially reducing the required kick strength and helping to mitigate the associated systematic uncertainties, as discussed in the following section.}

The main technical challenges of this approach lie in the detailed design and stabilization of the magnetic field.  
Because the storage region is composed of discrete sector magnets, the field uniformity and symmetry in the edge-transition regions must be precisely controlled. The angle of the magnet edge determines the focusing strength ($Q_h$, $Q_v$) and therefore requires a precise stabilization.
Moreover, the strong field gradients near the magnet edges render conventional NMR probes ineffective, as field inhomogeneities and line broadening can obscure the resonance signal.
\rev{Calibration of the orbit-averaged magnetic field may therefore require complementary techniques. Traditional NMR measurements could, for example, be supplemented by electron paramagnetic resonance (EPR) or fluxmeter measurements to constrain the field in the high-gradient regions.}
%One possibility is to combine traditional NMR with electron paramagnetic resonance (EPR) or fluxmeter measurements, which can remain sensitive in high-gradient regions.
A more innovative method follows Farley’s original idea of using a \emph{polarized proton beam} as an in-ring co-magnetometer. Protons with the same momentum as the stored muons are injected onto the corresponding closed orbit, and their spin-precession frequency is measured with a polarimeter based on spin-dependent elastic scattering from a thin internal target. With detectors placed symmetrically on the left and right sides of the target, the measured counting asymmetry is

\begin{equation}
  \epsilon(t)
  =
  \frac{N_L(t)-\alpha N_R(t)}
       {N_L(t)+\alpha N_R(t)}
  \simeq
  A_N P_0
  \cos\!\left(\omega^m_{a,p}t+\phi_0\right),
  \label{eq:polarimeter_asymmetry}
\end{equation}
where $A_N$ is the analyzing power, $P_0$ is the proton-beam polarization, $\phi_0$ is the initial spin phase, and $\alpha$ accounts for the relative
acceptance and efficiency of the two detectors. Fitting this oscillation
directly determines the measured proton anomalous-precession frequency
$\omega^m_{a,p}$. This method intrinsically provides an orbit-averaged field and avoids the absolute-calibration procedure and diamagnetic-shielding corrections associated with water-based NMR probes.
Moreover, retractable polarimeter stations could be installed at more than one location, for example near the boundaries between adjacent sector magnets. The full-turn spin evolution determines the closed-orbit-averaged magnetic field, while the phase advances measured between adjacent stations provide spatially resolved information on the integrated field between the stations, potentially enabling a further reduction in systematic uncertainty.

%The absolute polarization and analyzing power primarily affect the oscillation amplitude and therefore do not need to be known with the same precision as the frequency.
%Rather than determining the magnetic field from a local probe measurement, the polarimeter measures the proton spin-precession frequency and thereby determines the magnetic field naturally averaged over the entire closed orbit and  avoids the absolute calibration and diamagnetic-shielding corrections associated with water-based NMR probes.

%Retractable polarimeter stations could be installed at several azimuthal  locations, for example near the boundaries of the segmented magnets. The full-ring spin evolution determines the closed-orbit-averaged magnetic field, while the phase advances between adjacent stations provide additional information on the field contribution from each individual magnet.

The performance of the proton co-magnetometer depends critically on the control and characterization of the proton-beam phase space \revv{and on achieving a proton spin-coherence time sufficiently long compared with the frequency-fit interval.} We therefore envisage a \textit{pencil-like proton beam} with a small transverse size and angular divergence, low transverse emittance, and a narrow momentum distribution \revv{to limit the spin-tune spread}. Such a beam could be prepared through beam-line matching and momentum and transverse collimation. The associated reduction in intensity is acceptable because only a modest proton rate is required and protons do not decay during storage. Beam-position monitors or dedicated proton trackers would measure the beam centroid and profile, allowing the proton field measurements to be explicitly reweighted to the stored-muon distribution.

The field and effective magnetic boundaries would be monitored using complementary thermal, mechanical, magnetic, and beam-based diagnostics. Temperatures, magnet currents, and cooling conditions would be logged continuously, while displacement and tilt sensors and magnet-end fiducials referenced to a laser-tracker network would monitor mechanical motion. Fixed probes would track relative field drifts where the gradients permit their operation, supplemented by periodic proton-beam scans through the high-gradient edge regions. The effective focusing strength would also be monitored in situ through the measured horizontal and vertical tunes and the coherent-betatron-oscillation (CBO) frequency. Precision field mapping and continuous drift monitoring have been demonstrated in the Fermilab Muon $g\!-\!2$~\cite{Gioiosa2024Magnetometry,Corrodi2020FieldMapping}. The slow-control system would distinguish short and long timescale drifts using independent reference channels, following the general monitoring strategy adopted for the Fermilab laser calibration system~\cite{Anastasi2019LaserMonitoring}. 
%Quantitative sensitivity estimates are presented in Sec.~\ref{sec:sensitivity}.

%a ppb-level
%However, it requires a calibration beam with narrow momentum spread and long-term stability, and its spin-precession frequency must be measured at the ppb level.
%\missing{A full simulation of the injection process, magnetic-field design, and particle tracking will be required to firmly establish the feasibility of this concept.}

It is worth noting that potential synergies exist with several related projects. The proposed Electron-Ion Collider in China (EicC) also requires polarized proton beams and high-precision polarimetry~\cite{Anderle:2021wcy,Liang:2025owx}, \rev{while a non-destructive proton polarimeter has recently been proposed for storage-ring applications~\cite{Kim:2026dsl}}. In addition, the neutrino-factory-oriented nuSTORM project adopts a sector-magnet geometry~\cite{Franklin:2025muw}. These programmes may therefore provide opportunities for shared technical development.

%It is worth noting that another project under development, the proposed Electron-Ion Collider in China (EICC)~\cite{Anderle:2021wcy, Liang:2025owx}, will also rely on a polarized proton beam and the advancement of high-precision polarimeter techniques. The proton co-magnetometer proposed here could therefore be directly synergized with the EICC program, offering mutual benefits in both technical development. We also note that the design philosophy of the neutrino-factory–oriented nuSTORM project~\cite{Franklin:2025muw} shares strong similarities with the sector magnet concept here, suggesting potential synergies in future development. Also a recent proposal for the non-destructive proton polarimeter~\cite{Kim:2026dsl}.

\subsection{Concept B: Hybrid Weak Focusing Upgrades to the Fermilab Scheme}
\label{subsec:designB}

\begin{figure}[ht]
\centering
\includegraphics[width=0.98\linewidth]{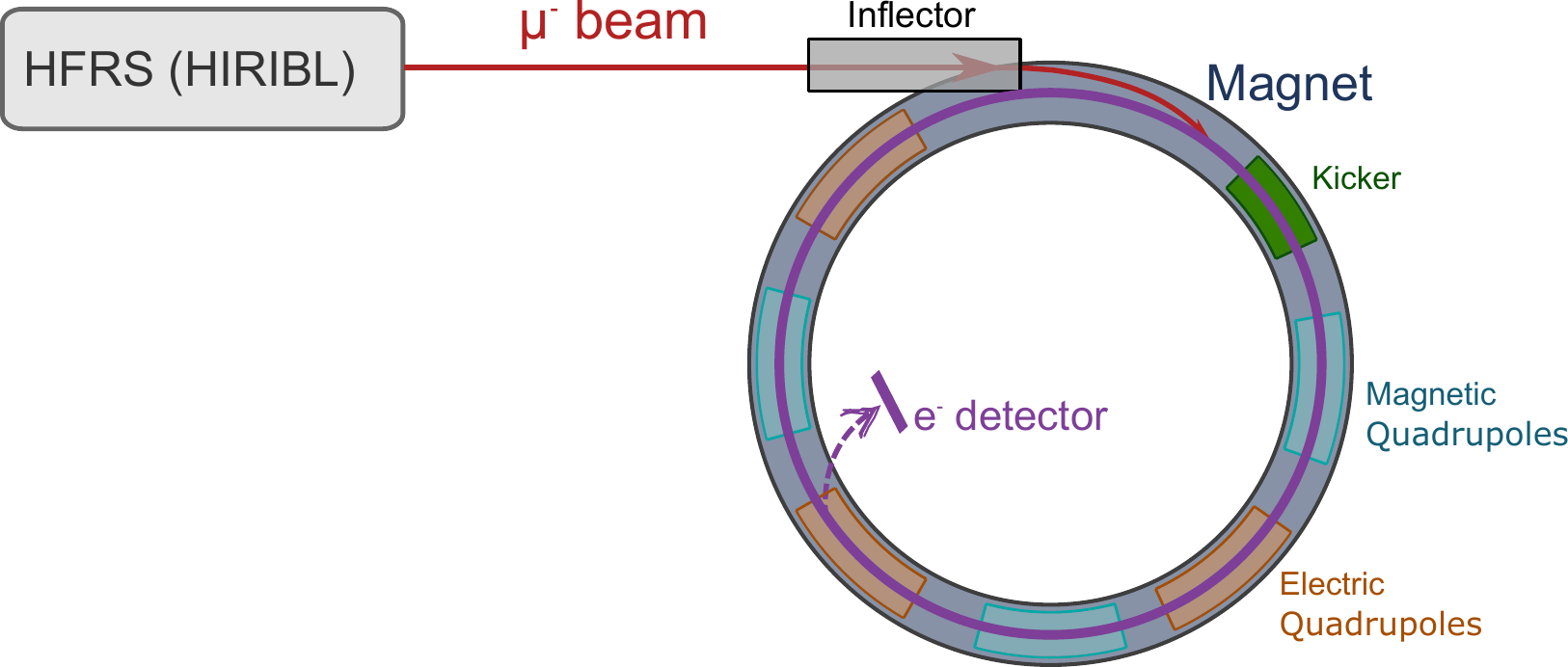}
\caption{\label{fig:designB}Schematic design of Concept~B: hybrid weak focusing using both magnetic and electric quadrupoles.}
\end{figure}

Concept~B, illustrated in Figure~\ref{fig:designB}, extends the philosophy of the Fermilab Muon~$g\!-\!2$ experiment by introducing a hybrid weak-focusing lattice that employs both electric and magnetic quadrupoles.  
In the conventional configuration, 
starting from Eq.~\eqref{eq:wa_general} and introducing a realistic stored-beam distribution with a normalized momentum deviation $\delta\equiv\frac{\Delta P}{p_{\rm mag}}$, the precession frequency may be written as
$\omega_a \simeq \omega_{a0}\bigl(1 + C_e \bigr)$,
where $\omega_{a0}=-\frac{q}{m}a_{\mu}B$ and the dispersive correction arising from electric quadrupoles is
\begin{equation}
C_e^{E\text{-focus}} \simeq 
-\frac{\beta^2}{a_\mu}
\left(a_\mu - \frac{1}{\gamma^2 - 1}\right)
\frac{n_E}{1 - n_E}\,\delta .
\end{equation}
This term vanishes at the magic momentum, but operating at that momentum
%However, the magic-momentum condition 
tightly constrains the stored-muon lifetime and thus the achievable statistical reach of the experiment.

The hybrid-focusing approach, first
\rev{proposed and studied}
in Ref.~\cite{Kim:2021pbo}, provides a mechanism to remove this constraint.  
By superimposing magnetic quadrupole gradients on the conventional electrostatic focusing field, the electric focusing index $n_E$ in the expression above is replaced by the combined focusing strength $n = n_E + n_B$, and magnetic focusing contributes an additional dispersive term
$C_e^{B\text{-focus}} \simeq -\frac{n}{1 - n}\,\delta$ .
The total correction to $\omega_a$ is therefore
\begin{equation}
C_e = C_e^{E\text{-focus}} + C_e^{B\text{-focus}} .
\end{equation}
One can tune the relative electric and magnetic focusing indices ($n_E$ and $n_B$) such that the combined contribution to the dispersive offset to $\omega_a$ cancels for a chosen, generally \emph{non-magic}, momentum.  
This “effective-magic” condition can be expressed approximately as
\begin{equation}
\frac{n_B}{n_E} = - \frac{\beta^2}{a_\mu} \left(a_\mu - \frac{1}{\gamma^2 - 1}\right),
\end{equation}
which generalizes the usual magic-momentum relation to include both electric and magnetic focusing components. The corresponding values of $n_B/n_E$ for representative muon momenta are shown in Fig.~6 of Ref.~\cite{Kim:2021pbo}.

For example, at $p_\mu = 5~\mathrm{GeV}/c$, one can achieve the cancellation by choosing $n_E = 0.10$ and $n_B = -0.062$.  
These indices can be scaled while preserving their ratio, allowing optimization according to storage efficiency, achievable quadrupole gradients, or beam-dynamics constraints such as betatron tunes.

Beyond enabling higher-momentum operation, the hybrid configuration introduces additional tunable parameters for optimizing field uniformity and exploring systematic dependencies \rev{by varying $n_E$ and $n_B$}.  
It thus serves as a powerful intermediate step between the present Fermilab ring and more radical next-generation storage-ring geometries, preserving the proven experimental framework while expanding its kinematic and operational flexibility. %Further simulation studies will be carried out to finalize the detailed design of this concept.
%\rev{The present tracking studies in Ref.~\cite{Kim:2021pbo} are limited to idealized single-particle trajectories. Further simulations will extend this treatment to realistic beam distributions and three-dimensional field maps in order to quantify the storage efficiency, the residual dispersive correction, and the associated magnetic-field and beam-dynamics uncertainties.}

%\rev{A practical implementation of magnetic quadrupoles in this Concept would employ conventional independently powered quadrupole windings or pole-face correction coils distributed around the ring. The required gradients are modest and can be tuned during commissioning to achieve the desired ratio $n_B/n_E$. Therefore, while detailed engineering optimization and field-quality studies remain part of the later design phase, we do not currently identify the realization of the magnetic quadrupole system itself as a major technical feasibility risk.}

\rev{
A practical ring would replace the continuous-gradient idealization of Ref.~\cite{Kim:2021pbo} with an azimuthally symmetric set of discrete magnetic-focusing sections. The magnetic-quadrupole component could be generated by independently powered quadrupole windings or pole-face correction coils distributed around the ring. For the field convention adopted here, the required gradient
%satisfies
%$
%  \left|G_B\right|
%  \equiv
%  \left|\frac{\partial B_y}{\partial x}\right|
%  =
%  \frac{|n_B|B_0}{R_0},
%$
%and 
is modest for the representative values of $n_B$ considered above. The coil currents would provide reproducible control of the gradient, while the effective ring-averaged value of $n_B$ would be determined from magnetic-field maps and beam-based tune measurements.

The generation of the quadrupole field therefore relies on mature accelerator-magnet technology. The specific challenge in Concept B is its integration into a precision storage field: fringe fields, higher multipoles and alignment must be characterized, and the resulting three-dimensional field must be included consistently in the field average and in the beam and spin-dynamics corrections. These requirements form part of the dedicated engineering and field-characterization programme for the later design phase.
}

\section{Projected Sensitivity and Precision Reach}
\label{sec:sensitivity}

%\vspace{0.5em}
\paragraph{Statistical considerations.}  
The statistical precision of the measurement is estimated using
Eq.~\eqref{eq:sensitivity}.
For the HIAF baseline (Phase-I), we adopt a representative muon momentum of approximately $3~\mathrm{GeV}/c$, within the expected range of $2\text{--}4~\mathrm{GeV}/c$, corresponding to $\gamma\approx28.4$. \rev{Although the magnetic field inside the sector magnets is assumed to be $3~\mathrm{T}$, the field-free straight sections reduce the effective field averaged over one complete turn. Depending on the three- or four-sector configuration, the effective field is $B_{\rm eff}\simeq1.5\text{--}2.0~\mathrm{T}$. The projected HIAF event sample is estimated by scaling the Fermilab sample of $3.09\times10^{11}$ analyzed decay positrons according to the muon flux ratio listed in Table~\ref{tab:beam-comparison}. Assuming comparable effective data-taking periods and overall storage and detection efficiencies, this scaling gives approximately $2.5\times10^{11}$ analyzed decay electrons or positrons at HIAF. Taken together, these assumptions yield a projected statistical precision of approximately $0.10\text{--}0.14~\mathrm{ppm}$, based on a scaling from the final Fermilab statistical uncertainty of $0.114~\mathrm{ppm}$.} 
%The improvements discussed in this proposal, such as the higher injection efficiency by the novel sector magnet design (Concept A), could further enhance this precision, though the gain is expected to be modest. 

The upgraded HIAF-U (\rev{Phase-II)} configuration offers \rev{a more substantial improvement}. Taking a representative muon momentum of $15~\mathrm{GeV}/c$, within the expected range of $10\text{--}20~\mathrm{GeV}/c$, gives $\gamma\approx142$ and a dilated muon lifetime of approximately $312~\mu\mathrm{s}$. 
\rev{A direct scaling from the final Fermilab statistical uncertainty, with a storage field of $3\text{--}6~\mathrm{T}$ ($B_{\rm eff}\simeq2\text{--}4~\mathrm{T}$),
gives
approximately $0.002$--$0.01~\mathrm{ppm}$ for the projected
Phase-II statistical precision.
We nevertheless retain
$\sigma_{\rm stat}\lesssim0.02~\mathrm{ppm}$ as a conservative target, so that the event sample required to reach this sensitivity could be
accumulated within a relatively short data-taking period.} At this level, the measurement would become dominated by systematic uncertainties \rev{and would approach the precision limit imposed by the external constants}.

\rev{
Table~\ref{tab:statistical-assumptions} summarizes the key parameters entering these statistical-sensitivity estimates for Phase-I and Phase-II.
}

\begin{table*}[ht]

\color{black}
\arrayrulecolor{black}

\centering
\caption{\rev{Comparison of the key parameters entering the statistical-sensitivity projections. The HIAF Phase-I event sample assumes an effective data-taking period comparable to that of FNAL Runs~1--6, whereas the HIAF-U Phase-II samples assume shorter data-taking periods. The statistical sensitivities for HIAF are obtained by scaling from the FNAL result according to Eq.~\eqref{eq:sensitivity}. All projections assume the same selection in normalized decay-electron or positron energy and the same asymmetry weighting as at FNAL.}}
\label{tab:statistical-assumptions}
\begin{tabular}{p{5.5cm}p{3.0cm}p{3.0cm}p{3.0cm}}
\toprule
Parameter & FNAL (Runs~1--6) & HIAF Phase-I & HIAF-U (Phase-II) \\
\midrule
Representative muon momentum (GeV/$c$)
  & 3.094
  & 3.0 \; (2--4)
  & 15 \; (10--20) \\

Lorentz factor ($\gamma$)
  & 29.3
  & $\sim28.4$
  & $\sim142$ \\

Effective orbit-averaged field, $B_{\rm eff}$ (T)
  & 1.45
  & 1.5--2.0
  & 2--4 \\

Muon polarization ($P$)
  & $\simeq95\%$
  & 85--90\%
  & $\gtrsim90\%$ \\

Analyzed decay electrons/positrons ($N$)
  & $3.09\times10^{11}$
  & $\sim2.5\times10^{11}$
  & $\sim6\times10^{11}$--$6\times10^{12}$ \\

Effective decay asymmetry ($|A|$)
  & $\sim38\%$
  & $\sim38\%$ (assumed)
  & $\sim38\%$ (assumed) \\

Statistical sensitivity (ppm)
  & 0.114 (achieved)
  & $\sim0.10$--$0.14$ (projected)
  & $\lesssim0.02$ (projected) \\
\bottomrule
\end{tabular}
\end{table*}

\paragraph{Systematic considerations.}  

\rev{
We discuss the sources of systematic uncertainty separately for the two concepts. For each concept, we first identify the uncertainties that are closely analogous to those in the Fermilab muon $g\!-\!2$ experiment. We then examine how the differences in SR design and operating conditions may modify these uncertainties, including cases in which an established Fermilab correction is substantially reduced or eliminated. Finally, we introduce systematic effects that are specific to the novel features of each concept and therefore have no direct counterpart in a Fermilab-style SR.

In the Fermilab analysis, each major systematic category is further divided into several underlying contributions, whose detailed evaluations are reported in Refs.~\cite{Muong-2:2021vma,Muong-2:2021xzz,Muong-2:2021ovs,Muong-2:2024hpx,Muong-2:2026qnz}. 
The benchmarks adopted in the following estimate are taken from the final Fermilab Runs 4--6 report~\cite{Muong-2:2026qnz}. A complete bottom-up evaluation of every subcomponent including detailed engineering designs and detector simulations, is beyond the scope of the present work. Instead, we identify the dominant contributions within each category and assess how they are expected to change in the proposed concepts.

It is also important to note that several Fermilab uncertainties could be further reduced even within a similar experimental configuration. Potential improvements include reliable operation at the full design kicker strength, improved collimation and temperature control, and refined procedures for magnetic-field operation and stabilization with the magnet-power-supply feedback loop. As discussed in Ref.~\cite{Hertzog:2025ssc}, the advances in detector performance and analysis methods, could plausibly reduce the relevant Fermilab-style total systematic uncertainty from approximately $78~\mathrm{ppb}$ to $25~\mathrm{ppb}$. Several of the strategies discussed in that review are incorporated into the estimates presented below.

In order to provide a sufficient discussion including detector-related effects, we adopt a representative Fermilab detector configuration for both concepts. The baseline consists of Fermilab-style electromagnetic calorimeters distributed around the ring (24 stations in the Fermilab muon $g\!-\!2$ experiment) to measure the arrival times, energies, and spatial distributions of decay positrons. We assume modest improvements in geometrical acceptance, segmentation and crystal response. We also assume several tracking stations for precision measurements of the stored-muon profiles.
Fermilab employed trackers at two azimuthal locations. While comparable single-station performance is adopted as the baseline, broader azimuthal coverage is assumed to improve the reconstruction of the stored-muon distribution. For example, for Concept~A, one tracker station could be placed near each sector magnet, providing three or four independent tracker measurements.

}

\subsection{\rev{Systematic-Uncertainty Targets for Concept~A}}
%%%% === Concept A

For Concept~A, the magnetic field is determined using a circulating polarized-proton beam. This alternative field-calibration scheme leads to a correspondingly different expression for extracting $a_\mu$. The anomalous spin precession of the proton beam, as determined from the spin-phase evolution measured by the polarimeter, is governed by the BMT equation in the same manner as Eq.~\eqref{eq:wa_general}:

\begin{equation}
\begin{split}
\vec{\omega}_{a,p}
= -\,\frac{q_p}{m_p}\Biggl[
a_p\,\vec B
- a_p\frac{\gamma_p}{\gamma_p+1}
  (\vec\beta_p\cdot\vec B)\vec\beta_p
\\
-\left(
a_p-\frac{1}{\gamma_p^2-1}
\right)
\frac{\vec\beta_p\times\vec E}{c}
\Biggr],
\end{split}
\label{eq:wa_proton}
\end{equation}
where $a_p=(g_p-2)/2$ is the proton magnetic moment anomaly,
$q_p$ and $m_p$ are the proton charge and mass, and
$\gamma_p$ and $\vec\beta_p$ are its Lorentz factor and normalized
velocity, respectively. In this approach, the reference protons
circulate freely in the SR rather than being bound within
the material of an NMR probe. Consequently, the conventional
conversion from the shielded-proton precession frequency in a water
sample to the corresponding free-proton quantity is not required.

Assuming that \(\vec E=0\) and, ideally,
\(\vec\beta\cdot\vec B=0\), the latter two terms vanish, giving
\begin{equation}
  \omega_{a,p}
  =
  \frac{|q_p|}{m_p}a_p B.
  \label{eq:a_p}
\end{equation}
This relation is analogous to the corresponding expression for the
muon in Eq.~\eqref{eq:wa_pureB}. The effective magnetic field sampled
by the stored muons can therefore be expressed as
\begin{equation}
  \widetilde B
  =
  \frac{m_p}{|q_p|a_p}
  \left\langle\omega_{a,p}\times M\right\rangle,
\end{equation}
where, as defined previously, $M$ denotes the normalized stored-muon
distribution used to weight the measured magnetic field. Combining
this expression with the corresponding muon precession frequency gives anomalous magnetic moment of muon

\begin{equation}
  a_\mu
  =
  \frac{\omega_{a,\mu}}
       {\left\langle\omega_{a,p}\times M\right\rangle}
  a_p
  \frac{m_\mu}{m_p}
  \frac{|q_p|}{|q_\mu|}.
  \label{eq:a_mu_extraction_proton}
\end{equation}
Since $|q_p|=|q_\mu|=e$, the charge ratio is unity and will be omitted in the following.

%\missing{\sout{Eq.~\eqref{eq:a_mu_extraction_proton} differs substantially from the Fermilab extraction formula in Eq.~\eqref{eq:a_mu_extraction_fnal}, as the two approaches depend on different external constants. In both the Fermilab and J-PARC extraction approaches, the relevant combination involves $\mu'_{\rm p}(\Tr)/\muBohr$ and $m_\mu/m_e$, with an uncertainty at the level of approximately $22~\mathrm{ppb}$~\cite{mohr2024codatarecommendedvaluesfundamental}. In the present approach, this combination is replaced by $a_p m_\mu/m_p$. The proton anomaly $a_p$ is known with a relative uncertainty of approximately $0.45~\mathrm{ppb}$, while $m_\mu/m_p$ is also known to approximately $22~\mathrm{ppb}$~\cite{mohr2024codatarecommendedvaluesfundamental}. The total external-parameter uncertainty is therefore at a comparable level. This alternative dependence on external constants, together with the different field-calibration chain, provides a complementary and largely independent cross-check of the Fermilab result.}}

\revv{Although Eq.~\eqref{eq:a_mu_extraction_proton} appears different from the Fermilab extraction formula in Eq.~\eqref{eq:a_mu_extraction_fnal}, their external-parameter content can be compared by rewriting the factor in the Fermilab
expression as
\begin{equation}
\frac{\mu'_p(T_r)}{\mu_B}\frac{m_\mu}{m_e}
=
(1-\sigma'_p)(1+a_p)\frac{m_\mu}{m_p}.
\end{equation}
Here, $\mu'_p(T_r)=(1-\sigma'_p)\mu_p$, where $\mu_p$ is the
free-proton magnetic moment and $\sigma'_p$ is the proton magnetic-shielding constant for a spherical water sample at the reference temperature $T_r$. Combining the shielding relation with $\mu_p/\mu_N=1+a_p$ and $\mu_N/\mu_B=m_e/m_p$ yields the expression above, where $\mu_N= e\hbar/(2m_p)$ is the nuclear magneton.

%Using $\mu_p/\mu_N=1+a_p$ and $\mu_N/\mu_B=m_e/m_p$, where $\mu_N$ is the nuclear magneton and $a_p=(g_p-2)/2$ is the free-proton magnetic anomaly, gives the relation above.

The two extraction methods therefore share $m_\mu/m_p$ as the dominant source of external-parameter uncertainty. For Concept~A, the remaining factor $a_p$ is known to a relative precision of approximately $0.45~\mathrm{ppb}$~\cite{mohr2024codatarecommendedvaluesfundamental}. The free-proton co-magnetometer also avoids the proton magnetic-shielding correction $\sigma'_p$ associated with the water-NMR standard, whose present uncertainty contribution is approximately $4.1~\mathrm{ppb}$. Ongoing high-precision muonium-spectroscopy experiments, including MuSEUM at J-PARC~\cite{Kanda:2020mmc} and Mu-MASS at PSI~\cite{Crivelli:2018vfe}, are expected to improve the relevant mass-ratio determinations, potentially reducing the combined external-parameter uncertainty to approximately $10~\mathrm{ppb}$~\cite{MuSEUM:2025cmo}. We therefore adopt a prospective external-parameter allocation of approximately $10~\mathrm{ppb}$ for Concept~A (and likewise for Concept~B).}

The physical muon anomalous-precession frequency $\omega_{a,\mu}$ is obtained from the measured frequency $\omega^m_{a,\mu}$ after applying the pitch correction $C_p$, the muon-loss correction $C_{ml}$, the phase-acceptance correction $C_{pa}$, and the differential-decay correction $C_{dd}$. Additional corrections are required for the proton-based determination of the effective magnetic field. Accounting for these effects, Eq.~\eqref{eq:a_mu_extraction_proton} becomes:

\begin{equation}
  a_\mu \;=\;
  \frac{\omega^m_{a,\mu}(1+C_p+C_{ml}+C_{dd}+C_{pa})}{f_{\mu/p}\langle\omega^m_{a,p}\times M\rangle(1+B_{k}+B_{p}+B_{pl}+B_{pa})}
  a_p
  %\frac{\mu_p'}{\mu_B}
  \frac{m_\mu}{m_p} \, .
  \label{eq:a_mu_extraction_proton_w_correction}
\end{equation}

The electric-field correction $C_e$, which is one of the major beam-dynamics corrections in the Fermilab analysis, is absent because no applied electric field is used. The correction $B_q$ associated with the pulsed electrostatic quadrupoles in Fermilab is also absent. The kicker-transient correction $B_k$ remains necessary. The use of a proton beam to determine the magnetic field introduces several additional corrections associated with proton-beam dynamics and polarimetry. These include the proton pitch correction $B_p$, arising when $\vec\beta_p\cdot\vec B\neq0$, a possible proton-loss correction $B_{pl}$, and the polarimeter phase-acceptance correction $B_{pa}$. The transfer factor $f_{\mu/p}$ accounts more generally for differences between the effective fields sampled by the proton and muon ensembles. It is defined such that $f_{\mu/p}=1$ when the proton and muon beams sample identical  magnetic fields.

The projected systematic uncertainty targets for each category are discussed in the following subsections.

\subsubsection{Muon precession frequency $\omega^m_{a,\mu}$}
The systematic uncertainty associated with $\omega^m_{a,\mu}$ reached 30~ppb in the final Fermilab analysis and was dominated by: \textit{(i)} the modelling of coherent betatron oscillation (CBO), \textit{(ii)} the various gain responses of the calorimeters and \textit{(iii)} the pileup. 

For Concept~A, the time-independent magnetic focusing eliminates the tune evolution associated with the settling of the electrostatic quadrupoles. In addition, injection through a field-free straight section avoids the restrictive inflector aperture and is expected to require only a small corrective kicker deflection, thereby reducing the coherent centroid mismatch and the coupling between beam motion and the detected electron or positron spectrum. Additional tracking stations can directly constrain the CBO amplitude, phase advance, and decoherence envelope, which were among the major contributors to the CBO systematic uncertainty in the later Fermilab datasets.
As a quantitative benchmark, the Fermilab CBO uncertainty in the later datasets~\cite{Muong-2:2024hpx,Muong-2:2026qnz} contained contributions of approximately 16~ppb from the modelling of the decoherence envelope and 10~ppb from the time dependence of the CBO frequency. The time-independent focusing conditions in Concept~A should be particularly beneficial for the latter contribution, since they suppress time-dependent tune evolution and therefore reduce the complexity of modelling the CBO-frequency evolution. Prospective design allocations of 14~ppb and 5~ppb, respectively, would correspond to $ \sigma_{\rm CBO}
 \simeq
 \sqrt{(14~{\rm ppb})^2+(5~{\rm ppb})^2}
 \simeq 15~{\rm ppb},$
which we adopt as the design target.

The lower repetition rate at HIAF increases the number of stored muons per fill and hence the instantaneous rate in each detector channel. 
If $r(t)$ denotes the single-electron or positron rate, the absolute accidental-pair rate scales approximately as $r^{2}(t)\Delta t_{\rm eff}$, whereas its fractional contribution to the single-particle spectrum scales as $r(t)\Delta t_{\rm eff}$. Neglecting the oscillatory modulation, the single-particle rate has an envelope proportional to $\exp(-t/\tau_{\rm lab})$, where $\tau_{\rm lab}=\gamma\tau_\mu$, and the accidental-pair envelope decreases approximately as $\exp(-2t/\tau_{\rm lab})$. Following the data-driven approach developed at Fermilab~\cite{Girotti:2023fbg}, the pileup spectrum can be reconstructed and its residual effect constrained directly from data. Reducing the repetition rate from approximately 11--12~Hz at Fermilab to 3~Hz would increase the pileup uncertainty by a factor of about three. Scaling from the Fermilab uncertainty of 5--6~ppb therefore gives an estimate of approximately 15--18~ppb before any detector and reconstruction improvements. The corresponding fractional
unresolved-pair contamination could reach the
$\mathcal{O}(10^{-3}$--$10^{-2})$ level in the most highly occupied
channels near the beginning of the fit interval and would decrease
rapidly at later times. 

To translate these estimates into preliminary detector requirements, we use the demonstrated Fermilab calorimeter performance as a benchmark. The Fermilab system consists of 24 stations with 54 independently read out crystals per station and uses 800~MS/s waveform digitization. It demonstrated a single-shower timing resolution below 100~ps for energies above 1.8~GeV, full separation efficiency for showers more than approximately 5~ns apart, and dead-time-free continuous waveform digitization throughout the 700-$\mu$s fill~\cite{Muong-2:2019beb}. A preliminary local design criterion for HIAF is therefore
$r_{\rm ch}(t_{\rm fit})\Delta t_{\rm eff}\lesssim10^{-2}.$
For $\Delta t_{\rm eff}\lesssim5~\mathrm{ns}$, this corresponds to a rate below approximately 2~MHz in the most highly occupied channel at the beginning of the fit interval. Taking the lower end of the above scaling estimate and assuming pileup-reconstruction performance comparable to that achieved at Fermilab, we adopt a Phase-I design target of
$\sigma_{\rm PU}^{\rm I}(\omega_a)\lesssim15~\mathrm{ppb}.$

The segmentation required for pileup control would also reduce the per-channel occupancy and the associated rate-dependent gain variations. Building on the experience gained at Fermilab, independently stabilized readout channels, improved bias and temperature control, and continuous in-fill gain calibration provide a credible route toward further improvement. Relative to the combined gain-related uncertainty of approximately 15--25~ppb achieved at Fermilab, we therefore adopt a
Phase-I design target of
$\sigma_{\rm gain}^{\rm I}(\omega_a)\lesssim10~\mathrm{ppb}.$

For Phase-II, the repetition rate is expected to be comparable to that at Fermilab, while the dilated lifetime at $\gamma\simeq140$ is longer by a factor of approximately four to five. This reduces the instantaneous decay rate per stored muon, but the expected factor-of-10--50 increase in the average muon rate may partly or fully offset this advantage. Assuming that the increase in delivered flux translates into a comparable increase in the stored-muon rate, and for otherwise unchanged detector acceptance and segmentation, the early-time per-channel rate could range from approximately one-half to three times that of Phase-I. A direct linear extrapolation from the adopted Phase-I pileup target would therefore give an upper benchmark of $\mathcal{O}(40~\mathrm{ppb})$ before any Phase-II-specific detector and reconstruction improvements.

Finer segmentation, improved two-pulse resolution, waveform-level multi-pulse fitting, spatial clustering, and data-driven pileup reconstruction could plausibly reduce this unmitigated estimate. The resulting reduction in per-channel occupancy, together with improved channel stabilization and continuous in-fill calibration, could also maintain the gain-related uncertainty near the Phase-I level despite the higher average muon rate. Moreover, the longer Phase-II development timescale would allow the detector granularity and readout architecture to be optimized using simulated local rate distributions. We therefore adopt the provisional Phase-II design targets
$\sigma_{\rm PU}^{\rm II}(\omega_a)\lesssim20~\mathrm{ppb},$ and
$\sigma_{\rm gain}^{\rm II}(\omega_a)\lesssim10~\mathrm{ppb}.$

For the Phase-I benchmark, combining the three  contributions discussed above in quadrature gives a projected uncertainty of approximately $24~\mathrm{ppb}$ on $\omega^m_{a,\mu}$ for Concept~A. For Phase-II, replacing the Phase-I pileup allocation of $15~\mathrm{ppb}$ with the provisional $20~\mathrm{ppb}$ gives a combined  uncertainty of approximately $27~\mathrm{ppb}$.

\subsubsection{Muon beam-dynamics corrections}

%, however, remains because vertically focused muons do not move exactly perpendicular to the magnetic field. The final

The absence of an electric field in Concept~A eliminates the electric-field correction ($C_e$), whose uncertainty was 27~ppb in the final Fermilab analysis. 
% The pitch correction $C_p$ originates from the vertical betatron motion, which makes $\beta\cdot B\neq0$. 
% \missing{In the Fermilab lattice, the electrostatic quadrupoles provide the vertical focusing and therefore enter the this dynamics.} 
The pitch correction $C_p$ arises from the coupling of vertical betatron motion to the spin dynamics, including the oscillating radial component of the spin-precession vector. In the Fermilab lattice, the vertical betatron motion is generated by electrostatic-quadrupole focusing, whereas in Concept~A it is governed primarily by edge focusing.
Fermilab assigned an uncertainty of approximately $9~\mathrm{ppb}$ to $C_p$, of which approximately $8~\mathrm{ppb}$ arose from tracker hardware and vertical-coordinate reconstruction, corresponding to an uncertainty of approximately $0.3~\mathrm{mm}$ in the reconstructed vertical beam width. Concept~A envisages tracking instrumentation comparable to, or more extensive than, that employed at Fermilab. Improved tracker alignment and resolution, and dedicated beam-profile measurements therefore provide a plausible route to $\sigma_{C_p}\lesssim9~\mathrm{ppb}$. 
We note that the more complex fringe fields associated with edge focusing may modify the magnitude of the pitch correction itself, but this does not necessarily imply a larger systematic uncertainty. The uncertainty is instead governed by how accurately the vertical phase-space distribution and the relevant edge-field maps can be measured and modelled. The possible impact of edge-focusing field on $C_p$ is discussed in greater detail in Sec.~\ref{subsubsec:edge_field_uniform}.

%A dedicated spin-tracking model incorporating the realistic edge-focused optics will therefore be required, with a level of precision comparable to that achieved in the Fermilab beam-dynamics modelling.

The uncertainty of phase-acceptance correction, $C_{pa}$, of $15~\mathrm{ppb}$ in the final Fermilab analysis was dominated by two comparable contributions of approximately $10~\mathrm{ppb}$ each, associated with the tracker/CBO reconstruction and the simulated decay phase maps, respectively. The calorimeter acceptance and asymmetry maps contributed only at the sub-ppb level. For Concept~A, we adopt a target of
$\sigma(C_{pa})\lesssim10~\mathrm{ppb}$. In quadrature, this would require
the two dominant contributions to be reduced from approximately
$10~\mathrm{ppb}$ to $6\text{--}7~\mathrm{ppb}$ each. Such an improvement
appears plausible with improved azimuthal tracking coverage and
data-driven validation of the simulated phase maps and inter-station
beam-transport model. Moreover, the time-independent focusing lattice of
Concept~A avoids the time-dependent effects from EQs that contributed to
the evolution of the transverse beam distribution at Fermilab.

The differential-decay correction arises because higher-momentum muons have slightly
longer dilated lifetimes, causing the ensemble-averaged momentum
and spin phase $\phi_0$ to evolve during a fill. To leading order, the correction can be written as
$
C_{dd}
\simeq
\frac{\sigma_\delta^2}
{\omega_a\gamma_0\tau_\mu}
\frac{d\phi_0}{d\delta},
%\label{eq:Cdd_relative_momentum}
$
where $\delta\equiv(p-p_0)/p_0$ is the fractional momentum
deviation from the reference momentum $p_0$,
$\sigma_\delta^2$ is its variance, and $\gamma_0$ is the
Lorentz factor corresponding to $p_0$.
The correction was first included in the Fermilab Run-2/3 analysis and was subsequently examined extensively in the final analysis, which assigned an uncertainty of approximately $27~\mathrm{ppb}$.

A shorter injected bunch and improved temporal overlap between the kicker pulse and the bunch would suppress the injection-induced correlation and improve the determination of the phase--momentum correlation. Tighter control of the stored-beam momentum spread and dedicated beam diagnostics, such as the MiniSciFi system employed during the final Fermilab running period, would further constrain the multidimensional injection distribution $(\tau,\delta,x,x',y,y',\phi_0)$, where $\tau$ denotes the relative arrival time within the injected bunch. A recent review likewise notes that improvements informed by the Fermilab experience could substantially reduce the corresponding uncertainty ~\cite{Hertzog:2025ssc}. These considerations motivate a Phase-I design target of
$\sigma^{\rm I}(C_{dd})\lesssim15~\mathrm{ppb}.$

For Phase-II, the lifetime-driven momentum evolution is explicitly suppressed by $1/\gamma_0$. The increase from $\gamma_0\simeq28$ in Phase-I to $\gamma_0\simeq140$ in Phase-II therefore reduces the underlying differential-decay effect by approximately a factor of five for otherwise comparable momentum distributions and phase--momentum slopes. In addition, if the accepted absolute momentum bite does not  increase proportionally with $p_0$, the fractional width $\sigma_\delta=\sigma_p/p_0$ decreases, providing a further quadratic suppression. Provided that the accepted momentum distribution and $d\phi_0/d\delta$ remain well constrained, we therefore adopt a provisional Phase-II design target of
$\sigma^{\rm II}(C_{dd})\lesssim10~\mathrm{ppb}$
subject to validation through dedicated injection, beam-diagnostics, and spin-tracking studies.

Finally, $C_{ml}$ is determined by actual momentum-dependent muon losses and is not intrinsically increased by pileup or hadron contamination: at the expected level of less than one stored hadron per fill, the hadronic component makes a negligible contribution to the calorimeter pile-up relative to the $\mathcal{O}(10^{3})$ detected muon-decay events per fill. Prompt hadronic signals occur predominantly before the nominal fit start time and can be rejected using timing and energy selections. With a dedicated loss-monitoring system and data-driven accidental-background subtraction, a conservative target of $\sigma(C_{ml})<3~\mathrm{ppb}$ is appropriate.

\subsubsection{Proton precession frequency $\omega^m_{a,p}$ and corrections}
\label{subsubsec:omega_p}

In the final Fermilab Muon $g\!-\!2$ analysis, the uncertainty of the muon-weighted magnetic field was $48~\mathrm{ppb}$. The intrinsic absolute calibration, the uncertainty inflation from cross-checks with independent absolute probes, the probe-environment corrections, and the transfer to the trolley probes contributed $\simeq34~\mathrm{ppb}$ and they are absent in this concept. A second contribution of comparable size arose from spatial field mapping, temporal interpolation, and muon weighting. The Fermilab muon-weighting uncertainty was dominated by closed-orbit distortions associated with electrostatic-quadrupole misalignment, which contributes about 7~ppb, whereas the direct tracker acceptance, resolution, and alignment terms were generally at or below the ppb level. Additional tracking stations, improved alignment and reconstruction, and the absence of electrostatic-quadrupole closed-orbit distortions therefore motivate a target of $20~\mathrm{ppb}$

Residual coherent betatron motion, detector-gain variations, and pileup in the proton polarimeter can bias the extraction of $\omega^m_{a,p}$. The parameter $\alpha$ in Eq.~\eqref{eq:polarimeter_asymmetry} accounts for the relative acceptance and efficiency of the two polarimeter arms. Again, due to the pencil beam, the detector acceptance and the pileup effects should be greatly reduced. Simultaneous monitoring of the two detector arms, segmented low-occupancy readout, and dedicated gain and beam-motion measurements should constrain these effects to a design uncertainty of $8\text{--}10~\mathrm{ppb}$ for the polarimeter extraction of $\omega^m_{a,p}$.

\revv{Maintaining proton spin coherence over the frequency-fit interval is also required for a ppb-level extraction of $\omega^m_{a,p}$. Momentum spread and lattice nonlinearities produce a spread in the proton spin tune that can damp the polarimeter-asymmetry oscillation. Dedicated proton spin-tracking studies, potentially complemented by future measurements of the proton momentum distribution and the asymmetry-decoherence envelope, will be required to establish the allowable momentum spread and associated lattice tolerances.}

The electrostatic-quadrupole transient present at Fermilab is absent and therefore $B_q=0$. However, the kicker transient correction $B_k$ still remains. In the Fermilab analysis, the measured transient was parametrized as $\Delta B(t)=\Delta B(t_0)e^{-(t-t_0)/\tau_k},$ giving the approximate frequency shift~\cite{Muong-2:2021ovs}
\begin{equation}
\frac{\Delta\omega_a}{\omega_a}
\simeq
\frac{\Delta B(t_0)}{B_{\rm ref}}
f_{\rm azi}f_{\rm trans}
\left(\frac{\tau_k}{\tau_k+\gamma\tau_\mu}\right)^2,
\label{eq:kicker_transient_shift}
\end{equation}
where $f_{\rm azi}$ is the azimuthal fraction subtended by the kickers and $f_{\rm trans}$ the muon-weighted transverse average. 
The final Fermilab analysis assigned a total kicker-transient uncertainty of approximately 20--24~ppb across the Run-4/5/6 datasets~\cite{Muong-2:2026qnz}, for which we use $22~\mathrm{ppb}$ as a representative value. The dominant contribution, approximately 14--17~ppb, arose from extrapolating the measured transient across the storage aperture.
We assume comparable residual transient amplitudes and waveforms per unit azimuth. This assumption is motivated by the required main kicker pulse being of the same order as at Fermilab, namely tens of mT over approximately 100~ns, although the residual transient will ultimately depend on the detailed kicker and vacuum-chamber geometry. The corresponding uncertainty scales as
\begin{equation}
\sigma(B_k)
\simeq
\sigma^{\rm FNAL}(B_k)
\frac{\Delta B(t_0)}{\Delta B^{\rm FNAL}(t_0)}
\frac{1.45~{\rm T}}{B_{\rm eff}}
\frac{f_{\rm azi}}{f_{\rm azi}^{\rm FNAL}}
\frac{f_{\rm trans}}{f_{\rm trans}^{\rm FNAL}}
R_\tau ,
\label{eq:Bk_scaling_conceptA}
\end{equation}
where $R_\tau
=
\left(
\frac{\tau_k+64~\mu{\rm s}}
     {\tau_k+62~\mu{\rm s}}
\right)^2
\simeq1.03 
$ uses the representative Fermilab
kicker-transient decay timescale $\tau_k\simeq50$--$100~\mu{\rm s}$.
Taking
$\frac{\Delta B(t_0)}{\Delta B^{\rm FNAL}(t_0)}
\simeq
\frac{f_{\rm azi}}{f_{\rm azi}^{\rm FNAL}}
\simeq
\frac{f_{\rm trans}}{f_{\rm trans}^{\rm FNAL}}
\simeq1$
and $B_{\rm eff}=1.5$--$2.0~{\rm T}$ gives the direct scaling benchmark
$\sigma(B_k)\simeq16\text{--}22~\mathrm{ppb}$
before any geometry-specific improvement.

The transient amplitude, decay constant, and spatial dependence of the Concept~A kicker will be measured using Faraday-rotation magnetometry on a prototype straight section, following the Fermilab programme and including mock-up-based spatial modelling. The additional eddy-current coupling to the adjacent sector-magnet edges, which is specific to this geometry, will be studied within the same framework and incorporated into spin-tracking simulations. Because this programme directly addresses the aperture extrapolation that dominated the Fermilab uncertainty, we adopt a design target of
$\sigma(B_k)\lesssim10~\mathrm{ppb}.$

Several proton-specific beam-dynamics corrections should also be evaluated, including the proton pitch correction $B_p$, proton loss $B_{pl}$, and the phase-acceptance correction of the polarimeter $B_{pa}$. The low transverse emittance of the pencil proton beam suppresses pitch angle and acceptance variations, while precise injection matching and beam-position monitoring constrain coherent centroid motion. We therefore expect $B_p$ and $B_{pa}$ to be subdominant. The proton losses can be monitored directly and are not expected to produce an uncertainty larger than the corresponding muon-loss contribution. We therefore assign a design uncertainty of $3~\mathrm{ppb}$ to $B_{pl}$ and, allowing for residual contributions from $B_p$ and $B_{pa}$, conservatively assign a combined uncertainty of $4\text{--}6~\mathrm{ppb}$ to these proton beam-dynamics corrections.

Finally, we define $f_{\mu/p}$ to account for any residual difference between the field averages sampled by the proton and muon ensembles after the explicit phase-space reweighting. This factor potentially includes imperfect transverse and momentum matching, species-dependent pitch and fringe-field sampling and differences in the kicker histories. We conservatively assign a target uncertainty of $15~\mathrm{ppb}$ to $f_{\mu/p}$, guided by the approximately $10~\mathrm{ppb}$ trolley-to-fixed-probe tying uncertainty in the Fermilab analysis. The quantitative basis for this allocation is examined below through
simulations.
 
%with $u(\omega_{a,\mu})\simeq20\text{--}25~\mathrm{ppb}$ and a beam-dynamics uncertainty of approximately $24~\mathrm{ppb}$ gives
%\begin{equation}
%u_{\rm syst}^{\rm exp}(a_\mu)\simeq\sqrt{u^2(\omega_{a,\mu})+u^2_{\rm BD}+u^2_{\rm field}}\simeq42\text{--}47~\mathrm{ppb}.
%\label{eq:concept_total_systematic}
%\end{equation}

\rev{
\subsubsection{Edge-Focusing Tolerances and Field Uniformity}
\label{subsubsec:edge_field_uniform}

Since the detailed edge-focusing magnet and fringe-field design requires dedicated development, a final specification of the fabrication tolerances and field uniformity is beyond the scope of the present study. We instead present two preliminary sensitivity studies via tracking and field simulations: (i) the response of the beam dynamics to the variations of the edge angle, such as possible fabrication and alignment errors; and (ii) the impact of magnetic-field nonuniformity and residual muon--proton sampling mismatch on the transfer from the proton-measured field average to the muon-weighted field average.

As discussed above, the determination of several muon beam-dynamics corrections, including $C_{pa}$ and $C_{dd}$, is governed primarily by detector performance and injection conditions, such as the kicker--bunch overlap, rather than by the field. Among these corrections, $C_p$ is the one most directly related to edge focusing, as the vertical focusing in Concept~A is provided by the edge fields rather than by EQs. We therefore use $C_p$ to quantify the sensitivity to the effective edge angle. For the proton beam, the corresponding pitch correction $B_p$ is expected to be small because the tightly collimated, low-emittance beam has a small vertical angular spread.

In the first-order hard-edge approximation, the focusing strength at each magnet edge is $K_e=\tan\theta_f/\rho$, where $\theta_f$ is the angle between the reference trajectory and the normal to the effective magnetic boundary at the crossing point, and $\rho$ is the reference bending radius. We denote by $y$ the vertical displacement from the reference orbit and by $y'=dy/ds$ the vertical trajectory slope, where $s$ is the longitudinal coordinate along the reference orbit. The subscripts ``in'' and ``out'' denote quantities immediately before and after the effective magnetic boundary, respectively. In the thin-lens approximation, $y_{\mathrm{out}}=y_{\mathrm{in}}$, whereas the trajectory slope receives the edge-focusing kick

\begin{equation}
\label{eq:edgekick}
y'_{\mathrm{out}}=y'_{\mathrm{in}}-\frac{\tan\theta_f}{\rho}\,y_{\mathrm{in}}=y'_{\mathrm{in}}-K_e y_{\mathrm{in}} .
\end{equation}

A small coherent variation of the edge angle changes the focusing strength according to
$\frac{\Delta K_e}{K_e}
\simeq
\frac{\Delta\theta_f}
{\sin\theta_f\cos\theta_f}.$
For the nominal value $\theta_f=10^\circ$, this gives
$
\frac{\Delta K_e}{K_e}
\simeq
0.585\%
\left(\frac{\Delta\theta_f}{1~\mathrm{mrad}}\right).$
Such a variation modifies both the horizontal and vertical tunes and the stored vertical angular distribution. Figure~\ref{fig:edge_tracking} shows illustrative single-particle trajectories for relative momentum offsets $\delta_p=\Delta p/p$ within $\pm2\%$ and coherent effective-edge-angle errors within $\pm10~\mathrm{mrad}$ in the horizontal and vertical planes, respectively.

\begin{figure}[htbp]
    \centering
    \includegraphics[width=0.99\linewidth]{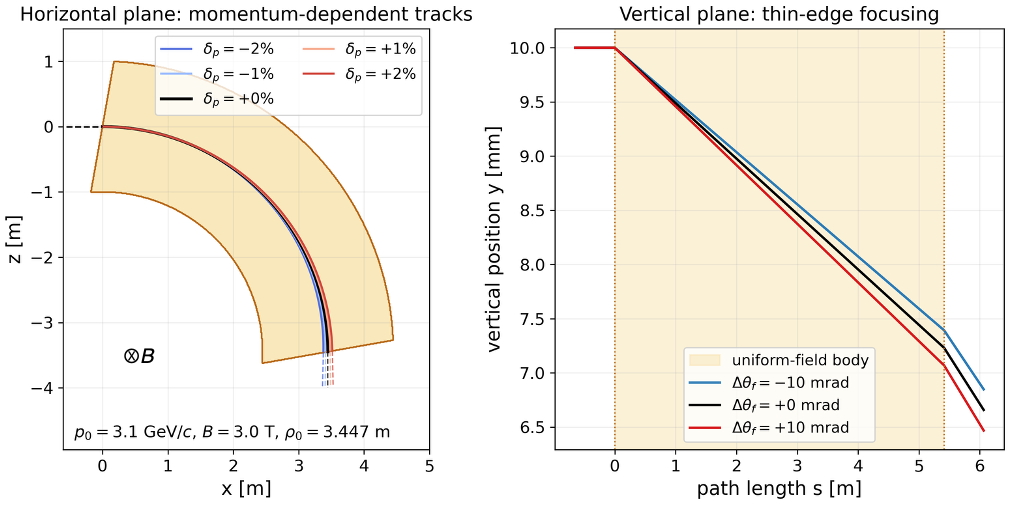}
    \caption{\rev{Single-particle tracking through a sector magnet. The left panel shows the horizontal trajectories for relative momentum offsets within $\pm2\%$, while the right panel shows the vertical trajectories for coherent effective-edge-angle errors within $\pm10~\mathrm{mrad}$.}}
    \label{fig:edge_tracking}
\end{figure}

To leading order, the pitch correction can be written as

\begin{equation}
\label{eq:C_p}
C_p
\simeq
\frac{1}{2}\left\langle\psi^2\right\rangle_B
\simeq
\frac{1}{2}\left\langle y'^2\right\rangle_B,
\end{equation}
where the average is taken over the stored-muon ensemble and the magnetic bending regions. 
%Although the edge kick is determined directly by $K_e$, the resulting change in $C_p$ also depends on the injected vertical phase-space distribution, its matching to the ring, and its subsequent evolution. 
We evaluate the relation between $\Delta\theta_f$ and
$\Delta C_p$ using a linear storage transport simulation.
For the four-sector lattice with $p=3.1~\mathrm{GeV}/c$, $B=3~\mathrm{T}$, and $\theta_f=10^\circ$, the nominal full-ring tunes are $Q_x=1.0198$ and $Q_y=0.5428$. Across the scanned range $\lvert\Delta\theta_f\rvert\leq10~\mathrm{mrad}$, the shifts in $Q_x$ and $Q_y$ remain below approximately $0.0071$ and $0.0166$, respectively. Therefore the lattice remains linearly stable.
Figure~\ref{fig:edge_sensitivity} shows the resulting sensitivity of $C_p$ to the edge-angle error. Adopting the conservative Fermilab-like benchmark
$C_{p,0}=170~\mathrm{ppb}$, corresponding to an rms pitch of $0.583~\mathrm{mrad}$, the linear transport simulation gives
$
\frac{dC_p}{d\theta_f}
\simeq0.573~\mathrm{ppb\,mrad}^{-1}.
$

\begin{figure}[htbp]
    \centering
    \includegraphics[width=0.88\linewidth]{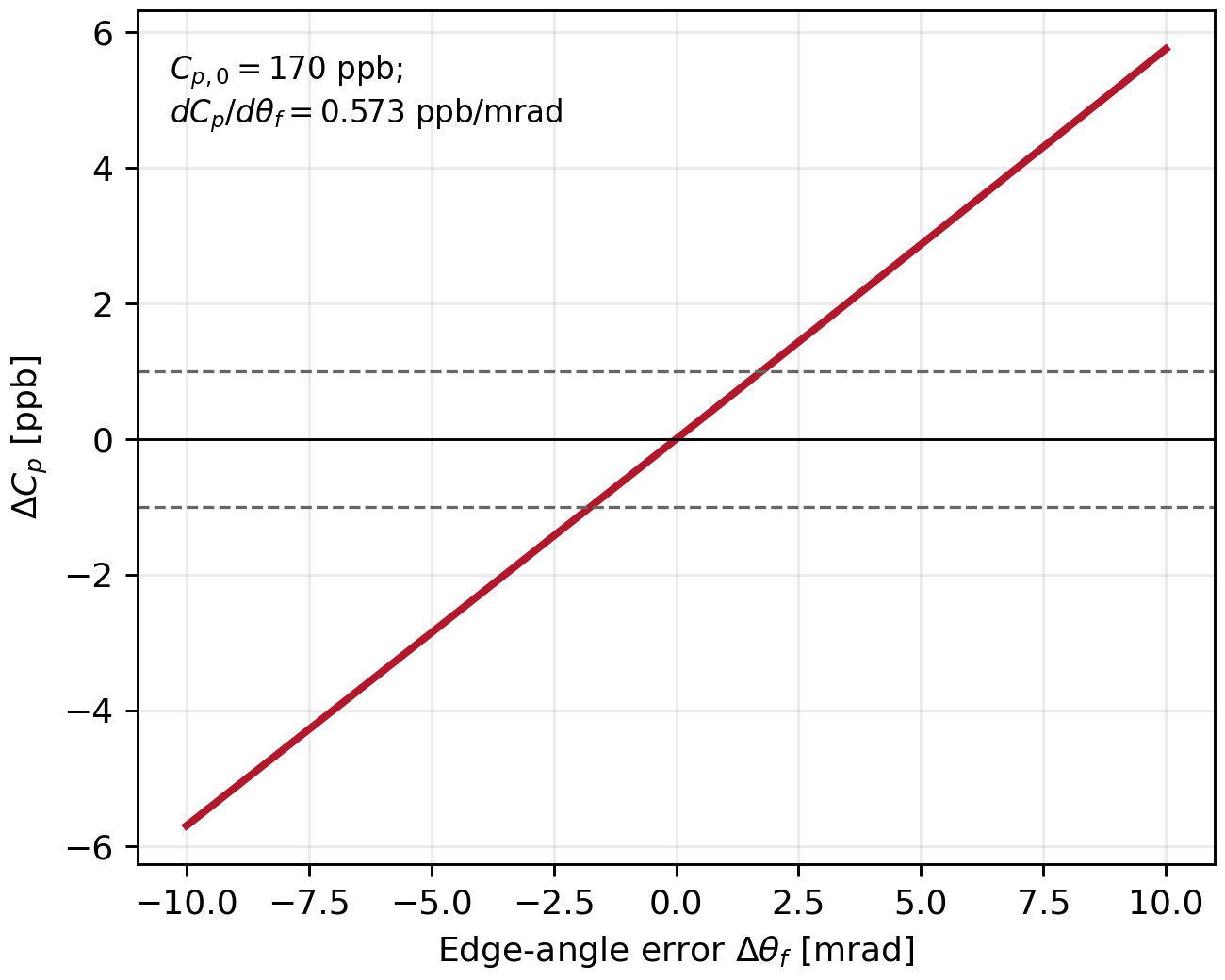}
    \caption{\rev{Change in the pitch correction $C_p$ as a function of the edge-angle error $\Delta\theta_f$. The calculation uses the benchmark $C_{p,0}=170~\mathrm{ppb}$, corresponding to an rms pitch of $0.583~\mathrm{mrad}$. The dashed lines indicate the $\pm1~\mathrm{ppb}$ allocation.}}
    \label{fig:edge_sensitivity}
\end{figure}
Therefore, allocating $1~\mathrm{ppb}$ to edge-angle-induced changes in $C_p$ requires the edge-angle error to be stable or independently constrained to approximately $1.7~\mathrm{mrad}$ in this model. This is a tolerance on the effective magnetic boundary rather than directly on the machined pole-face angle. Static fabrication errors can be characterized using three-dimensional field maps and potentially reduced by pole-end shimming or incorporated into the beam-dynamics model.

%The tune shifts also provide an in-situ monitor of the integrated focusing strength.  In the same model, the CBO frequency changes by approximately $7.7~\mathrm{kHz}$ per mrad of coherent edge-angle variation.  Synthetic decay spectra confirm that a five-parameter fit that neglects CBO is not adequate, whereas a fit including the injected tune-constrained CBO terms recovers the input $\omega_a$ in a noise-free closure test.  This study does not yet include nonlinear fringe fields, detector acceptance, or time-dependent decoherence, which will be required for a realistic uncertainty assignment.

To study the sensitivity of the proton-to-muon field transfer to magnetic-field nonuniformity and residual sampling mismatch, we define the fractional field-transfer residual as

\begin{equation}
r_B \equiv
\frac{
\widehat{\langle B\rangle}_{p}
-
\langle B\rangle_\mu
}{B_0},
\label{eq:field_transfer_residual}
\end{equation}
where $\widehat{\langle B\rangle}_{p}$ is the field average
reconstructed with the pencil proton-beam and reweighted to the muon distribution, while $\langle B\rangle_\mu$ is the true muon-weighted field average in the simulation. 
This residual is directly related to the field-transfer factor introduced in
Eq.~\eqref{eq:a_mu_extraction_proton_w_correction} through

\begin{equation}
f_{\mu/p}
\equiv
\frac{\langle B\rangle_\mu}
{\widehat{\langle B\rangle}_{p}},
\end{equation}
such that, for $\lvert r_B\rvert\ll1$, $f_{\mu/p}-1 \simeq -r_B$. The ensemble-rms residual, $R_B\equiv10^9\sqrt{\langle r_B^2\rangle}$ in ppb, therefore directly quantifies the scale relevant to the uncertainty allocation for $f_{\mu/p}$.

To illustrate the relevant tolerances, we generated an ensemble of 300 transverse, azimuthally averaged random-field maps with a $15~\mathrm{mm}$ correlation length. The rms field nonuniformity was varied between $0.1$ and $1~\mathrm{ppm}$. Representative Gaussian muon beam widths of $\sigma_x=10~\mathrm{mm}$ and $\sigma_y=10~\mathrm{mm}$ were assumed.
The proton beam was treated as a well-collimated pencil beam that provides a localized measurement of the magnetic field. It was assumed to be steered across the transverse aperture in $5~\mathrm{mm}$ steps, which denotes the spacing between successive sampling positions. The resulting field samples were interpolated and reweighted to the muon distribution in simulation.

For a $1~\mathrm{ppm}$-rms field nonuniformity and zero registration error, the finite $5~\mathrm{mm}$ sampling produces an rms interpolation residual of approximately $8.3~\mathrm{ppb}$. Figure~\ref{fig:field_transfer} shows the combined dependence of $R_B$ on the field nonuniformity and the residual muon--proton mean-orbit mismatch. The white contour denotes the $15~\mathrm{ppb}$ design uncertainty allocated to $f_{\mu/p}$. This contour corresponds to a residual mean-orbit mismatch of approximately $0.38~\mathrm{mm}$ for a $1~\mathrm{ppm}$-rms field nonuniformity. Conversely, for a fixed mismatch of $0.50~\mathrm{mm}$, the same criterion corresponds to a field nonuniformity of approximately $0.81~\mathrm{ppm}$ rms.

\begin{figure}[htbp]
    \centering
    \includegraphics[width=0.88\linewidth]
    {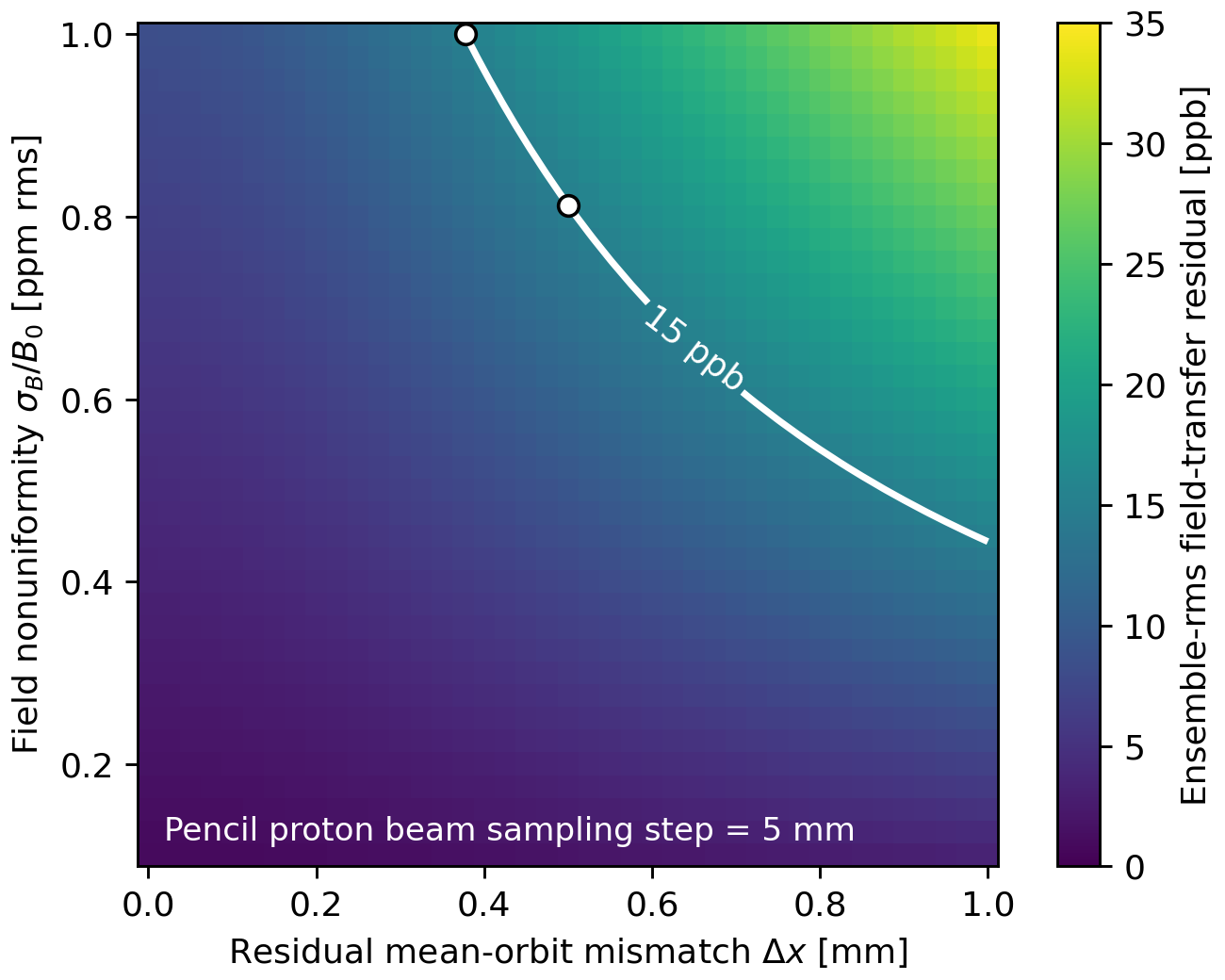}
    \caption{\rev{Ensemble-rms muon--proton field-transfer residual $R_B$ as a function of the residual mean-orbit mismatch and the rms field nonuniformity. The proton beam is sampled across the transverse aperture in $5~\mathrm{mm}$ steps. The white curve denotes the $15~\mathrm{ppb}$ design uncertainty allocated to the transfer factor $f_{\mu/p}$.}}
    \label{fig:field_transfer}
\end{figure}

Although the present ensemble study is preliminary, it indicates that an rms field nonuniformity of approximately $1~\mathrm{ppm}$, together with a residual muon--proton mean-orbit mismatch below approximately $0.4~\mathrm{mm}$, provides a useful design benchmark consistent with the $15~\mathrm{ppb}$ uncertainty allocation for $f_{\mu/p}$. We regard these as technically achievable design requirements in view of the monitoring and alignment strategy outlined in Sec.~\ref{subsec:designA}. Their final values should be established during the later R\&D stage with three-dimensional prototype field maps and more realistic proton and muon distributions. \revv{These field maps and beam distributions, together with dedicated proton spin-tracking studies, will then enable a quantitative determination of the proton spin-coherence time and the corresponding requirements on the proton momentum spread and lattice nonlinearities discussed in Sec.~\ref{subsubsec:omega_p}, forming a further key component of the co-magnetometer R\&D programme.}

Combining all contributions from muon precession frequency, muon beam dynamics corrections, proton precession frequency, field corrections, and the \revv{$10~\mathrm{ppb}$} uncertainty from external constants in quadrature, we obtain a total projected systematic uncertainty of approximately \revv{$46~\mathrm{ppb}$}. The corresponding uncertainty budget and a comparison between the final Fermilab analysis and Concept~A are presented in Table~\ref{tab:sys_compare}.

}

\subsection{\rev{Systematic-Uncertainty Targets for Concept~B}}
%%%% Concept B

Concept~B retains the continuous-ring geometry, conventional NMR field-calibration chain, electrostatic-focusing component, kicker injection, and decay-positron detection strategy employed at Fermilab.
Consequently, the extraction of $a_\mu$ follows the same structure as the Fermilab analysis. Expanding Eq.~\eqref{eq:a_mu_extraction_fnal} to display the individual corrections gives

\begin{equation}
  a_\mu \;=\;
  \frac{\omega^m_{a,\mu}(1+C_e+C_p+C_{ml}+C_{dd}+C_{pa})}{\langle\omega^m_{p}(\Tr)\times M\rangle(1+B_{k}+B_{q})}
  \frac{\mu'_\text{p}(\Tr)}{\muBohr} \frac{m_{\mu}}{m_\text{e}}.
  \label{eq:a_mu_extraction_ConceptB}
\end{equation}

This expression is identical in structure to that used in the final Fermilab analysis~\cite{Muong-2:2026qnz}. The determination of $\omega^m_{a,\mu}$ relies on essentially the same calorimeter, tracking, and time-spectrum analysis techniques at Fermilab, together with the potential improvements as discussed in Concept~A. We therefore adopt common design targets of $15~\mathrm{ppb}$ for CBO modelling and $10~\mathrm{ppb}$ for gain-related detector effects, together with pileup targets of $15~\mathrm{ppb}$ for Phase-I and $20~\mathrm{ppb}$ for Phase-II, giving combined targets of $24~\mathrm{ppb}$ and $27~\mathrm{ppb}$ for Phase-I and Phase-II, respectively.

The major Concept-B-specific improvement concerns $C_e$, which now denotes the net dispersive correction produced by both the electric and magnetic focusing components, $C_e=C_e^{E\text{-focus}}+C_e^{B\text{-focus}}$. At the effective-magic operating point, the leading momentum-dependent contributions from electric and magnetic focusing cancel. 
For a design momentum $p_0$ and the corresponding Lorentz factor
$\gamma_0$, the cancellation condition is
\begin{equation}
    \frac{n_B}{n_E}
    =
    -\frac{\beta_0^2}{a_\mu}
    \left(
        a_\mu-\frac{1}{\gamma_0^2-1}
    \right).
    \label{eq:effective_magic_condition}
\end{equation}
For a muon with momentum
$p=p_0(1+\delta)$, where
$\delta\equiv(p-p_0)/p_0$, an expansion about the design momentum gives
\begin{align}
    -\frac{\beta^2}{a_\mu}
    \left(
        a_\mu-\frac{1}{\gamma^2-1}
    \right)
    &\simeq
    -\frac{\beta_0^2}{a_\mu}
    \left(
        a_\mu-\frac{1}{\gamma_0^2-1}
        +\frac{2\delta}{\gamma_0^2-1}
    \right)
    \nonumber\\
    &=
    \frac{n_B}{n_E}
    -2\left(
        \beta_0^2+\frac{n_B}{n_E}
    \right)\delta .
\end{align}

The leading residual electric-field correction near the cancellation
point is therefore second order in the momentum deviation:
\begin{equation}
    C_e(p\simeq p_0)
    \simeq
    -2\,
    \frac{\beta_0^2 n_E+n_B}
         {1-n_B-n_E}\,
    \delta^2 .
    \label{eq:Ce_effective_magic_second_order}
\end{equation}
This expression is analogous to the higher-order correction at the
conventional magic momentum,
\begin{equation}
    C_e(p\simeq p_m)
    \simeq
    -2\beta_m^2
    \frac{n_E}{1-n_E}\,
    \delta^2 .
\end{equation}
As illustrated in Fig.~7 of Ref.~\cite{Kim:2021pbo}, the expected magnitude of $C_e$ around the Concept-B design momentum is of order $1~\mathrm{ppm}$, comparable to that in the Fermilab Muon $g\!-\!2$ experiment. Fermilab demonstrated that the uncertainty associated with such a correction can be controlled at the level of a few tens of ppb. Further reduction should be possible with tighter control of the stored-beam momentum distribution and weaker time--momentum correlations.

Moreover, in Concept~B, independent variation
of $n_E$ and $n_B$ provides a direct experimental means of locating
the cancellation point and testing the residual momentum dependence.
Together with precise measurements of the electric and magnetic field
indices and improved spin-tracking techniques building on those
developed at Fermilab, these considerations support a design target of
$\sigma(C_e)\lesssim20~\mathrm{ppb}.$

The magnetic-quadrupole field also enters the pitch correction and
higher-order spin--orbit terms. Its inclusion may therefore modify the central value of $C_p$. Based on the Fermilab analysis experience, however, its uncertainty remains governed mainly by the measurement and modelling technique for the stored vertical phase space, provided that the three-dimensional magnetic field is sufficiently well characterized. A similar argument applies to $C_{pa}$, $C_{dd}$, and $C_{ml}$: although their central values may depend on the modified lattice and stored-beam distribution, the additional magnetic-quadrupole fields are not expected to introduce qualitatively new dominant sources of uncertainty. We therefore adopt targets similar to those for Concept~A:
$\sigma(C_p)\lesssim9~\mathrm{ppb}$,
$\sigma(C_{pa})\lesssim10~\mathrm{ppb}$,
$\sigma(C_{dd})\lesssim15~\mathrm{ppb}$ for Phase~I, improving to
$\lesssim10~\mathrm{ppb}$ for Phase~II, and
$\sigma(C_{ml})\lesssim3~\mathrm{ppb}$.

The conventional Fermilab field-calibration chain---absolute calibration, transfer to the trolley probes, field mapping, temporal interpolation with the fixed probes, and muon weighting---can be retained without introducing a fundamentally new calibration principle. Nevertheless, there are several routes to improving the precision. For absolute calibration, ${}^{3}\mathrm{He}$ magnetometers could be employed as an independent calibration reference and, if technically feasible, incorporated directly into the trolley. Compared with the conventional water-based NMR standard, such probes are expected to have reduced sensitivity to field gradients, temperature variations, and sample-geometry-dependent corrections~\cite{Hertzog:2025ssc}. Together with redundant calibration probes, improved positioning reproducibility, and a temperature-stabilized calibration environment, these developments support a design allocation of approximately $25~\mathrm{ppb}$ for absolute and trolley-probe calibration.

The uncertainty associated with field mapping, temporal interpolation, and muon weighting could likewise be reduced by placing a denser array of fixed probes closer to the storage region, improving their cross-calibration with the trolley probes, and performing more frequent trolley surveys. Improved thermal stability of the experimental hall and magnetic-field feedback would further suppress temporal drifts.
Moreover, the better-collimated and more centrally stored beam envisaged for Concept~B would sample a smaller and more uniform region of the field. Increased tracker coverage, together with the improvements in tracker alignment and resolution discussed for Concept~A, would provide a more complete three-dimensional measurement of the stored-muon distribution, reducing the uncertainty in the muon-weighted field. 

The magnetic-quadrupole windings would remain energized during the trolley surveys, such that the NMR system maps the total storage field rather than treating the focusing field as a separate correction.
As in the Fermilab analysis, measurements from a transverse array of trolley probes can be decomposed at each azimuth into normal and skew multipole components. Measurements at several quadrupole-current settings, including current-off reference scans, would isolate the applied quadrupole response and test its reproducibility. 

NMR remains applicable in the presence of the planned gradient, although the field variation across the probe volume shortens the free-induction-decay signal and increases the sensitivity to probe positioning and the frequency-extraction procedure. Given the modest gradient envisaged and the established use of transverse probe arrays for multipole field mapping, these effects are not expected to impose a fundamental limitation on the field measurement. Complementary rotating-coil or calibrated Hall-probe surveys would constrain the transverse and fringe-field structure, while fixed probes placed at several transverse positions, together with precision current monitoring, would track temporal variations of both the dipole and quadrupole components. Finally, beam-based measurements of the betatron tunes and dedicated scans of $n_E$ and $n_B$ would provide an independent constraint on the integrated focusing strength and directly locate the effective-magic operating point. The resulting field map would then be weighted by the measured muon distribution in the same manner as the conventional Fermilab field map.
Together, we adopt a design allocation of
approximately $20~\mathrm{ppb}$ as the design target for field mapping, interpolation, and muon weighting. We also note that, in principle, the polarized-proton co-magnetometer could also be deployed for Concept~B, providing an independent cross-check of the NMR-based field determination.

The electrostatic-quadrupole transient $B_q$ remains relevant to Concept~B, but it could in principle be eliminated, rather than merely corrected, by operating the ESQs with sufficiently long voltage pulses or, ideally, a static voltage throughout the measurement period. This approach was not feasible at Fermilab for two main reasons. First, the capacitance of the pulser-switch system prevented the required high voltage from being sustained for the necessary timescale, which can extend to seconds. Second, selected ESQ plates were intentionally operated with time-dependent voltages during the initial scraping period, thereby displacing the stored beam toward the collimators, removing large amplitude muons, and reducing subsequent muon losses.
For Concept~B, these limitations could be avoided through improved upstream beam collimation and the development of a robust long-pulse or DC ESQ system. Together with a mechanically stable electrode structure and dedicated in-situ mapping of any residual transient fields, this could reduce the uncertainty from $20~\mathrm{ppb}$ at Fermilab to a conservative design target of $\sigma(B_q)\lesssim10~\mathrm{ppb},$
with the possibility of rendering this contribution negligible if fully static ESQ operation can be achieved.
Similarly, improved kicker construction and spatially resolved transient-field measurements, together with other kicker improvements discussed in Concept A motivate a similar target of $10~\mathrm{ppb}$ for $B_k$. 

%\missing{\sout{For the external parameters entering Eq.~\eqref{eq:a_mu_extraction_ConceptB}, ongoing high-precision muonium-spectroscopy experiments, including MuSEUM at J-PARC~\cite{Kanda:2020mmc} and Mu-MASS at PSI~\cite{Crivelli:2018vfe}, are expected to improve their determination, potentially reducing the associated combined uncertainty to approximately $10~\mathrm{ppb}$~\cite{MuSEUM:2025cmo}.}}
\revv{As discussed for Concept~A, we adopt the same prospective external-parameter allocation of approximately $10~\mathrm{ppb}$ for Concept~B.}
Combining this contribution in quadrature with the systematic uncertainties discussed above gives an overall systematic-uncertainty target of $\lesssim53~\mathrm{ppb}$.

Table~\ref{tab:sys_compare} summarizes the systematic-uncertainty targets for Concepts~A and B. For simplicity, the quoted totals use the more conservative of the Phase~I and Phase~II allocations. The phase-dependent differences in the pileup and differential-decay uncertainties have only a minor impact on the overall systematic budget.

\begin{table*}[htb]
\centering

\color{black}
\arrayrulecolor{black}

\caption{\rev{
Comparison of systematic uncertainties in the final Fermilab Runs~4--6 result and the indicative Phase-I and Phase-II design targets for Concepts~A and B. Where the two phases differ, the entries are shown as Phase-I/Phase-II. All entries are in ppb.
}}
\label{tab:sys_compare}

\begin{tabular}{lccc}
\toprule
Systematic source
& FNAL Runs~4--6
& Concept~A
& Concept~B \\
\midrule

\multicolumn{4}{l}{\textit{Muon precession-frequency extraction}}\\
CBO modelling
& 19--24 & $\leq15$ & $\leq15$ \\
Gain-related detector effects
& 15--25 & $\leq10$ & $\leq10$ \\
Pileup
& 5--6
& $\leq15$ (I) / $\leq20$ (II)
& $\leq15$ (I) / $\leq20$ (II) \\
Total $\omega^m_{a,\mu}$ systematic
& 30
& $\leq24$ (I) / $\leq27$ (II)
& $\leq24$ (I) / $\leq27$ (II) \\

\midrule
\multicolumn{4}{l}{\textit{Muon beam-dynamics corrections}}\\
Dispersive/electric-field correction $C_e$
& 27 & -- & $\leq20$ \\
Pitch correction $C_p$
& 9 & $\leq9$ & $\leq9$ \\
Muon-loss correction $C_{ml}$
& 2 & $\leq3$ & $\leq3$ \\
Phase-acceptance correction $C_{pa}$
& 15 & $\leq10$ & $\leq10$ \\
Differential-decay correction $C_{dd}$
& 27
& $\leq15$ (I) / $\leq10$ (II)
& $\leq15$ (I) / $\leq10$ (II) \\

\midrule
\multicolumn{4}{l}{\textit{Magnetic-field measurement and corrections}}\\
Absolute and trolley-probe calibration
& 34 & -- & $\leq25$ \\
Field mapping, interpolation, and muon weighting
& 34 & $\leq20$ & $\leq20$ \\
Electrostatic-quadrupole transient $B_q$
& 20 & -- & $\leq10$ \\
Kicker transient $B_k$
& 22 & $\leq10$ & $\leq10$ \\
Proton corrections $B_p$, $B_{pl}$, and $B_{pa}$
& -- & $\leq6$ & -- \\
Proton-frequency extraction $\omega^m_{a,p}$
& -- & $\leq10$ & -- \\
Muon--proton field-transfer factor $f_{\mu/p}$
& -- & $\leq15$ & -- \\

\midrule
\multicolumn{4}{l}{\textit{External parameters}}\\
$\mu'_{\rm p}(\Tr)/\muBohr$ and $m_\mu/m_e$
& 22 & -- & $\leq10$ \\
$a_p m_\mu/m_p$
& -- & \revv{$\leq10$} & -- \\

\midrule
Total experimental systematic
& 79 & \revv{$\leq46$} (both phases) & $\leq53$ (both phases) \\
\bottomrule
\end{tabular}
\end{table*}

\subsection{\rev{Additional Systematics Considerations for Phase-II}}
\rev{

Regarding Phase-II systematics, several additional R\&D studies will ultimately be required. For Concept~A, local sector fields approaching $6~\mathrm{T}$ would correspond to an orbit-averaged field of approximately $3$--$4~\mathrm{T}$, and hence to a radius of approximately $12.5$--$16.7~\mathrm{m}$ and a cyclotron period of $260$--$350~\mathrm{ns}$. A full-ring field of $6~\mathrm{T}$, as might instead be considered for Concept~B, would give $R\simeq8.3~\mathrm{m}$ and $T_c\simeq175~\mathrm{ns}$. A higher field can therefore prevent the ring dimensions from scaling directly with momentum, although the required high-field storage volume remains substantially more demanding than that at Fermilab. These timescales also show that a bunch width close to the prospective $100~\mathrm{ns}$ target, rather than the conservative $400~\mathrm{ns}$ baseline, would be necessary for Phase-II.
For field measurement in Concept~A, the larger circumference, higher local dipole fields, extended fringe-field regions, and the production and polarimetry of a same-rigidity proton beam at $10$--$20~\mathrm{GeV}/c$ introduce additional engineering and calibration challenges. For
Concept~B, the conventional NMR calibration chain can in principle be extended to higher field, with a correspondingly denser fixed-probe array and full-aperture trolley mapping.
Achieving and characterizing the required field uniformity over a large-aperture, multi-tesla ring at a relative accuracy comparable to that assumed for Phase-I has yet to be demonstrated.

At $\gamma\simeq142$ ($15~\mathrm{GeV}/c$), the decay electrons are more forward-collimated and extend to substantially higher energies.
The Phase-II detector would therefore require re-optimized azimuthal coverage, calorimeter geometry, and tracking acceptance. The longer measurement window would also impose more stringent requirements on detector gain and timing stability. 
Concept~A has no intrinsic high-energy kinematic restriction.
For Concept~B, Eq.~\eqref{eq:effective_magic_condition} gives $n_B/n_E\simeq-0.96$ at $15~\mathrm{GeV}/c$. The electric and magnetic focusing components therefore nearly cancel, leaving a relatively small net focusing index. A dedicated lattice study is required, as the existing proof-of-principle study at $5~\mathrm{GeV}/c$ cannot be extrapolated directly to this regime.

Nevertheless, several systematic effects scale favourably with energy. The differential-decay correction contains an explicit $1/\gamma$ suppression, and the fractional kicker-transient effect is also reduced at higher storage field. For a fixed number of stored muons per fill, the longer dilated lifetime also reduces the initial instantaneous decay rate and hence the severity of pileup. Accordingly, the Phase-II goal of $0.05~\mathrm{ppm}$ is conditional on achieving $\sigma_{\rm stat}\lesssim20~\mathrm{ppb}$ while controlling the total experimental systematic uncertainty at approximately the $50~\mathrm{ppb}$
level. A level of systematic control comparable to that targeted for Phase-I therefore provides an appropriate benchmark for Phase-II.

The higher available statistics would provide several interesting data-driven systematic controls. These include momentum and intensity scans, alternating $\mu^+$ and $\mu^-$ operation, collimation studies, and, for Concept~B, independent scans of $n_E$ and $n_B$ to locate the effective-magic point and vary the betatron tunes. Such measurements would directly test field weighting, detector acceptance, and momentum-dependent muon losses.

Finally, a comparison of the relevant experiments is summarized in Table~\ref{tab:experiment-summary}. Combining the statistical sensitivities in Table~\ref{tab:statistical-assumptions}, which are supported by HIAF beam simulations, with the systematic-uncertainty targets in Table~\ref{tab:sys_compare}, we project an overall precision target of $0.10~\mathrm{ppm}$ for Phase~I, while Phase~II could potentially reach $0.05~\mathrm{ppm}$.

}

\begin{table*}[t]
\centering
\color{black}
\arrayrulecolor{black}
\small
\setlength{\tabcolsep}{4pt}
\renewcommand{\arraystretch}{1.18}

\caption{\rev{Comparison of completed, planned, and proposed muon
$g\!-\!2$ experiments. The quoted precisions for E34 and
HIAF are projected targets.}}
\label{tab:experiment-summary}

\begin{tabular}{@{}L{2.2cm}L{2.45cm}L{2.45cm}L{3.2cm}L{6.4cm}@{}}
\toprule
 & \textbf{Brookhaven E821}
 & \textbf{Fermilab E989}
 & \textbf{J-PARC E34}
 & \textbf{HIAF (CANTON-$\mu$)} \\
\midrule

\textbf{Status}
& Completed
& Completed
& Under development
& Proposed \\

\textbf{Muon polarity}
& $\mu^+$ and $\mu^-$
& $\mu^+$
& $\mu^+$
& $\mu^+$ and $\mu^-$ \\

\textbf{Muon momentum}
& $3.1~\mathrm{GeV}/c$
& $3.1~\mathrm{GeV}/c$
& $300~\mathrm{MeV}/c$
& \makecell[tl]{$2$--$4~\mathrm{GeV}/c$ (Phase~I)\\
                $10$--$20~\mathrm{GeV}/c$ (Phase~II)} \\

\textbf{Storage and focusing}
& Continuous dipole with EQs
& Continuous dipole with EQs
& Compact solenoidal without EQ
& \makecell[tl]{Sector magnets with edge focusing (Concept~A)\\
                Full-ring hybrid weak focusing (Concept~B)} \\

\textbf{Magnetic-field determination}
& NMR trolley and fixed probes
& NMR trolley and fixed probes
& NMR probes
& \makecell[tl]{Polarized-proton co-magnetometer (Concept~A)\\
                NMR trolley and fixed probes (Concept~B)} \\

\textbf{Precision or design target}
& \makecell[tl]{$0.7~\mathrm{ppm}$ ($\mu^+$)\\
                $0.7~\mathrm{ppm}$ ($\mu^-$)}
& $0.127~\mathrm{ppm}$
& \makecell[tl]{$\sim0.45~\mathrm{ppm}$ (Phase~I)\\
                \rev{$\sim0.10~\mathrm{ppm}$ (Phase~II)}}
& \makecell[tl]{$\sim0.10~\mathrm{ppm}$ (Phase~I)\\
                $\sim0.05~\mathrm{ppm}$ (Phase~II)} \\

\bottomrule
\end{tabular}
\end{table*}

\section{Unique New Physics Potentials}
\label{sec:newphysics}
There have been numerous review papers discussing possible new physics interpretations of the muon $g\!-\!2$ anomaly. Comprehensive discussions can be found in the Muon $g\!-\!2$ Theory Initiative report~\cite{Aliberti:2025beg} and in dedicated reviews of beyond-the-Standard Model (BSM) scenarios~\cite{Athron:2025ets,DiLuzio:2021uty,Crivellin:2022gfu,Darme:2022yal,Coyle:2023nmi}.
In this section we highlight two aspects particularly relevant for the present proposal: 
the significance and energy scale of potential new physics implied by the current or future $g\!-\!2$ deviations from the SM, and 
the unique CPT sensitivity test that can be obtained from precision measurements of negative muons ($\mu^-$) in comparison with positive muons ($\mu^+$).
\paragraph{Significance and Energy Reach.}
Two representative Standard Model predictions lead to different levels of tension with the current world average. For the SM2020 prediction, the deviation is $\Delta a_\mu \equiv a_\mu^{\rm exp} - a_\mu^{\rm SM2020} = 2.574(454)\times10^{-9}$, corresponding to a significance of about $5.7\,\sigma$. In contrast, using the more recent SM2025 prediction, the deviation is reduced to $\Delta a_\mu = 0.344 (637)\times10^{-9}$, with a current significance of only $0.54\,\sigma$, which does not indicate any anomaly. An illustrative case comparing the BNL negative-muon measurement with SM2025 yields \rev{$\Delta a_\mu = 1.07(1.05) \times 10^{-9} \ (\sim 1\,\sigma)$}.
Steady improvements in the Standard Model prediction are anticipated, with a theoretical uncertainty at the level of $\sim 0.1\ \mathrm{ppm}$ representing a realistic near-term target, comparable to the precision already achieved by the Fermilab measurement.

\begin{figure}[ht]
\centering
\includegraphics[width=0.99\linewidth]{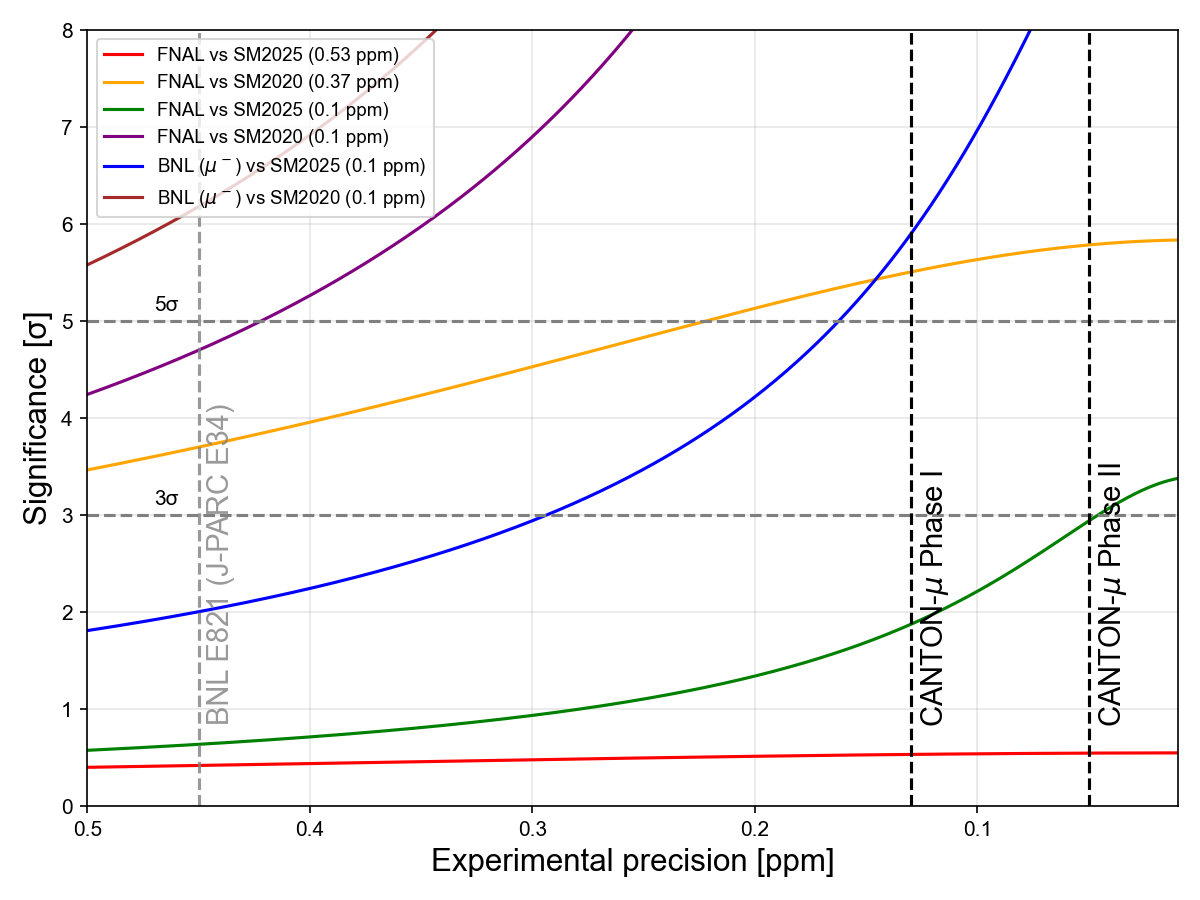}
\caption{
\label{fig:NP_scale}
$\Delta a_\mu$ significance as a function of experimental precision for several representative scenarios (Fermilab/BNL $\mu^-$ vs.\ SM2020/SM2025), with a projected theory uncertainty of 0.1 ppm. 
Phase-I can reach 5$\sigma$ for the experimental central value centered on BNL vs.\ SM2025, and Phase-II can reach $\sim$3$\sigma$ for FNAL vs.\ SM2025. 
}
\end{figure}

The interplay between experimental precision and statistical significance for several representative scenarios is illustrated in Fig.~\ref{fig:NP_scale}. At a precision comparable to the Fermilab result ($\sim 0.13\,\mathrm{ppm}$), corresponding to the CANTON Phase-I goal, a result consistent with the current BNL $\mu^-$ central value would yield a discrepancy exceeding $5\sigma$ relative to the SM2025 prediction (assuming a theoretical uncertainty of $0.1\,\mathrm{ppm}$). This provides a discovery-level test that is not accessible with existing $\mu^+$ data alone. At the subsequent Phase-II stage, 
\rev{if the conditional goal of $0.05~\mathrm{ppm}$ were
achieved,}
%with an improved experimental precision of $0.05\,\mathrm{ppm}$, 
even a measurement consistent with the current Fermilab central value would lead to a $\sim 2.9\,\sigma$ deviation with respect to the SM2025 prediction. For now, we must assume the measurements lie at the existing central values (FNAL or BNL); future experiments may update these central values. Taken together, these two regimes highlight the complementary discovery potential of the proposed program.

Now we discuss the corresponding new-physics energy scale $\Lambda$ that $\Delta a_\mu$ can probe, both at present and with future precision improvements. In a generic effective field theory framework, the contribution from new physics can be parametrized as

\begin{equation}
\Delta a_\mu \equiv a_\mu^{\rm exp} - a_\mu^{\rm SM} \sim C\,\frac{m_\mu^2}{\Lambda^2},
\end{equation}
where $m_\mu$ is the muon mass and $C$ is a dimensionless coefficient encoding the coupling strength $g$ and possible loop suppression or other enhancement factors. For tree-level contributions, one typically expects $C \sim g^2$, whereas loop-induced contributions generically give $C \sim g^2/(16\pi^2)$, as is common in renormalisable quantum field theories.

The implications for new physics depend strongly on the assumed size of $\Delta a_\mu$. Under the SM2020 prediction, the large deviation $\Delta a_\mu \sim \mathcal{O}(10^{-9})$ corresponds to a characteristic scale of $\Lambda_{\rm tree} \sim \mathcal{O}(5~\mathrm{TeV})$ for tree-level interactions, while loop-induced scenarios typically probe the sub-TeV to TeV range, depending on the coupling strength and possible enhancement mechanisms. This indicates relatively light new particles or sizeable couplings, whose parameter space is already tightly constrained by collider searches.
Robust lower bounds of around $100~\mathrm{GeV}$ on the masses have been established by LEP~\cite{L3:2001xsz,ALEPH:2001oot}, while the LHC further extends sensitivity to the TeV scale for weakly interacting particles and to multi-TeV scales for strongly interacting ones. 
In contrast, the much smaller deviation implied by the SM2025 prediction shifts the interpretation significantly. It probes higher effective energy scales, as smaller $\Delta a_\mu$ can be accommodated by heavier states.

The above discussion is based on scenarios without chiral enhancement, in which electroweak and chiral symmetry breaking are provided solely by the Standard Model Higgs vacuum expectation value and the muon Yukawa coupling. However, $a_\mu$ is intrinsically sensitive to chirality violation~\cite{Stockinger:2022ata, Czarnecki:2001pv, Athron:2025ets,Crivellin:2021rbq}, directly linking its magnitude to the muon mass and the underlying symmetry-breaking mechanism. Following the discussion of chiral enhancements in Ref.~\cite{Athron:2025ets}, this results in a modified scaling behaviour, which can be parametrized as
\begin{equation}
\label{eq:chiral_d_amu}
\Delta a_\mu \sim R_\chi \times \frac{c_L c_R}{16\pi^2}\ \frac{m_\mu^2}{\Lambda^2},
\end{equation}
where $c_L$ and $c_R$ denote the couplings to left- and right-handed muons respectively, and $R_\chi$ is a dimensionless enhancement factor that can satisfy $R_\chi \gg 1$.

A broad class of new-physics scenarios features strongly chirally enhanced contributions to $a_\mu$, which significantly increase the sensitivity to high mass scales. In such models, even very heavy states can generate observable effects, allowing $a_\mu$ to place important complementary constraints on the parameter space across a variety of scenarios. For example, models with radiative muon mass generation can yield $\Delta a_\mu \sim 10^{-9}$ for relevant mass scales of around 3~TeV. The maximum possible chiral enhancement in one-loop diagrams, with couplings near the perturbative unitarity limit, can generate $\Delta a_\mu \sim 10^{-9}$ even for mass scales up to 100~TeV, representing the largest plausible contribution for any given mass scale without violating perturbative unitarity~\cite{Capdevilla:2021rwo}. 

\begin{figure}
\includegraphics[width=0.99\linewidth]{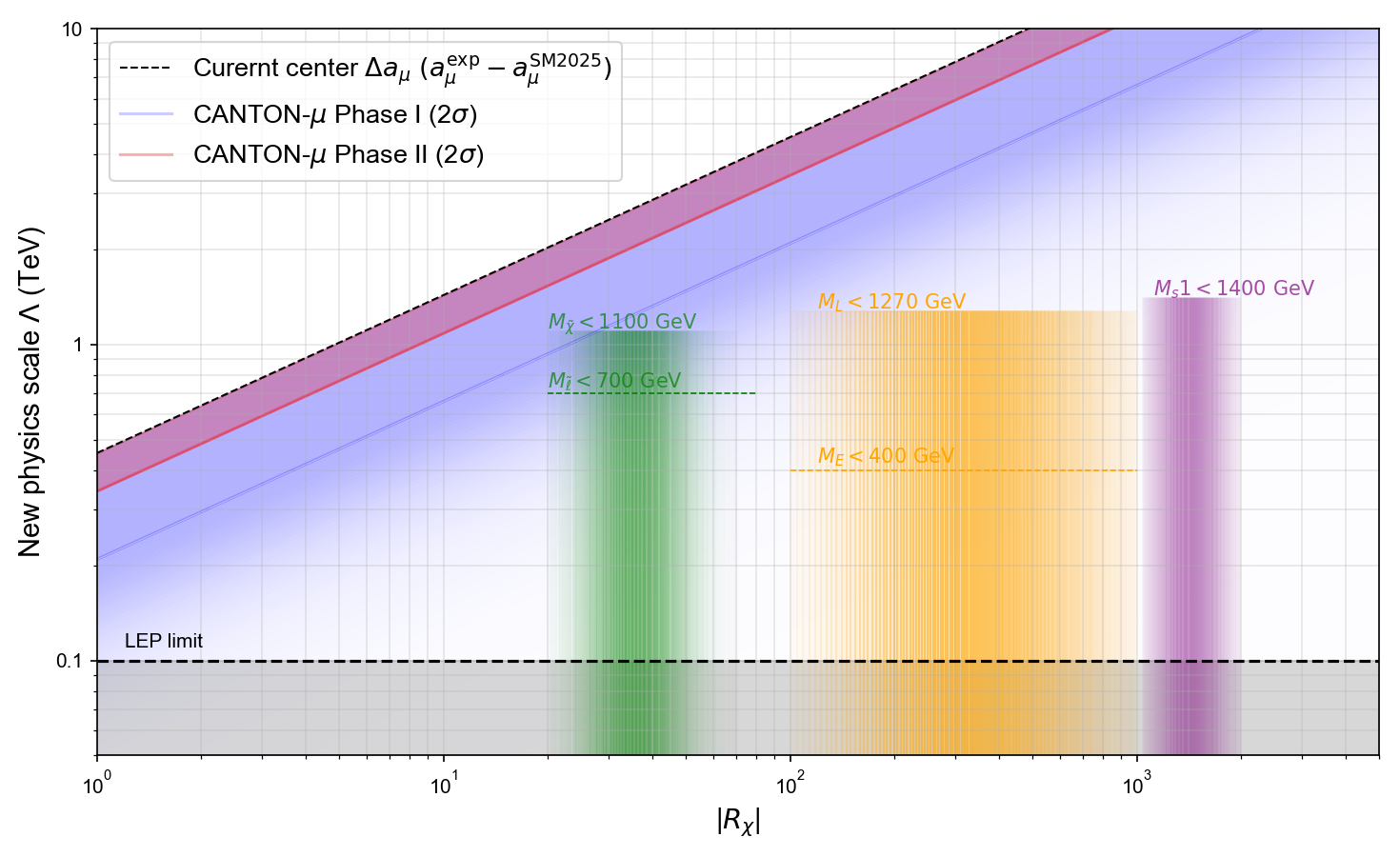}
\caption{
\label{fig:NP_scale2}
Mass exclusion limits from $\Delta a_\mu$ and direct collider searches as a function of the chiral enhancement factor $R_\chi$. The projected sensitivities of CANTON-$\mu$ Phase-I (blue) and Phase-II (red) are shown as $2\sigma$ lower limits on the new-physics mass scale, assuming $c_L, c_R \sim \mathcal{O}(1)$ in Eq.~\eqref{eq:chiral_d_amu} and centered at the current FNAL–SM2025 value. The current $2\sigma$ limit is not significant and is shown as a blue gradient. Representative LHC limits are also shown.
}
\end{figure}

Assuming the current experimental uncertainty, the lower limit on the relevant mass scale can be estimated. A conservative upper bound on $\Delta a_\mu = a_\mu^{\rm exp} - a_\mu^{\rm SM2025}$ is obtained from the $+2\,\sigma$ variation. Using the chirally enhanced scaling in Eq.\eqref{eq:chiral_d_amu}, this upper bound can be directly translated into a lower limit on the new-physics mass scale.
Figure~\ref{fig:NP_scale2} provides an illustrative comparison of mass exclusion limits from $a_\mu$ measurements and direct collider searches as a function of the chiral enhancement factor $R_\chi$, based on the compilation in Fig.~3.9 of Ref.~\cite{Athron:2025ets}. 
The projected capabilities of CANTON-$\mu$ Phase-I and Phase-II are shown by the solid blue and red lines, respectively, representing the corresponding $2\sigma$ lower limits on the new-physics mass scale derived from $\Delta a_\mu$, centered at the current FNAL–SM2025 value and assuming $c_L, c_R \sim \mathcal{O}(1)$ in Eq.~\eqref{eq:chiral_d_amu}. 
Phase-I corresponds approximately to the present FNAL precision. Depending on potential shifts in the central value, the exclusion curve could move to higher masses for smaller $\Delta a_\mu$, or approach the SM2020 scenario with enhanced sensitivity to lighter states for larger $\Delta a_\mu$. While the present $2\sigma$ bounds already compete with LHC exclusions, their statistical significance is low. 
With the improved precision of Phase-I and Phase-II, not only will the limits extend to higher masses, but the expected significance increases to $\sim 3\sigma$. 

Increased precision in $a_\mu$ measurements would enable $a_\mu$ to fully exceed the reach of current collider sensitivities. Several illustrative LHC limits are also shown, including mass reaches for charginos and sleptons (green) in the minimal supersymmetric standard model~\cite{Cox:2021nbo, ATLAS:2019lff}, vector-like leptons with SM-like doublet and singlet $M_L$ and $M_E$ (orange)~\cite{CMS:2024bni,ATLAS:2024mrr}, as well as scalar leptoquark types with chiral enhancement (purple)~\cite{Buchmuller:1986zs,ATLAS:2022wcu,CMS:2025iix,CMS:2023bdh,CMS:2024bnj}. For a comprehensive discussion of these new physics scenarios, see Ref.~\cite{Athron:2025ets}.
We also note that collider constraints are often highly model dependent, relying on specific production modes and kinematic signatures. In realistic scenarios 
%with multiple states or modified decay patterns, 
these limits can be significantly weakened. In this context, precision measurements of $a_\mu$ provide a powerful and complementary probe.

While a dedicated, model-specific analysis is beyond the scope of this proposal, we point out that $\Delta a_\mu$ can meaningfully constrain parameter spaces in various new physics scenarios. Currently, the theoretical predictions for $\Delta a_\mu$, by the SM2020 and SM2025 results, exhibit non-overlapping uncertainties, leaving vast regions of the relevant parameter space largely unconstrained. Future improvements in the precision of the $\Delta a_\mu$ measurement could therefore significantly clarify this landscape. 
Relevant illustrative results are shown in Ref.~\cite{Athron:2025ets}, including SMEFT studies (Fig.~2.5), dark-sector models such as light dark matter or dark photons (Figs.~5.2 and 5.4), ALP scenarios (Fig.~5.7), and SUSY models (Fig.~5.11), and more examples throughout the paper. In these cases, improved precision in $a_\mu$ measurements can narrow allowed parameter regions, impose stringent bounds, and help disentangle the interplay between $a_\mu$, LHC searches, and dark-matter constraints.

It should be emphasized that these considerations are highly scenario-dependent and the discussion here is illustrative rather than exhaustive, yet they already demonstrate the unique and powerful role of precise measurement of muon anomaly in probing and constraining new physics.

%In the Standard-Model Extension (SME) framework, the dominant framework for systematic CPT and Lorentz invariance tests, the CPT- and Lorentz-violating signatures for muons are encoded in a set of coefficients \cite{Bluhm:1999dx,Kostelecky:2008ts}. The corresponding coefficients have not been constrained by high-precision data for over two decades due to the absence of a modern $\mu^-$ $g\!-\!2$ measurement.

\paragraph{CPT tests from $\mu^+/\mu^-$ comparison.}
The most precise $\mu^-$ measurement is the Brookhaven E821 result~\cite{muon_g-2_collaboration_measurement_2004}, with an uncertainty of 0.7~ppm, about five times larger than the current $\mu^+$ world average.
This disparity leaves open the possibility of subtle differences between charge-conjugate states, providing a uniquely sensitive test of CPT symmetry~\cite{Quinn:2019ppv}. Within the Standard-Model Extension (SME) framework~\cite{Bluhm:1999dx}, differences between $\mu^+$ and $\mu^-$ anomaly frequencies are parametrized by coefficients such as $b_\kappa$. 
The time-averaged difference at a single site can be approximately expressed as
\begin{equation}
\Delta\omega_a \;=\; \langle\omega_a^{\mu^+}\rangle - \langle\omega_a^{\mu^-}\rangle
\;\approx\; \frac{4\,b_Z}{\gamma}\,\cos\chi \, ,
\end{equation}
where $\chi$ is the experimental colatitude, $b_Z$ is the projection of $b_\kappa$ along Earth's axis, and $\gamma$ is the Lorentz factor. 
The BNL E821 experiment reported $b_Z = (-1.0 \pm 1.1)\times 10^{-23}~\mathrm{GeV}$~\cite{Muong-2:2007ofc}, a limit that remains the most stringent to date. Since $b_Z$ can only be accessed through a combined $\mu^+$/$\mu^-$ analysis, improving this bound requires a new measurement of the negative-muon $g\!-\!2$ at comparable precision. 
The proposed HIAF facility, located at $23.1^\circ$N ($\chi \approx 67^\circ$), offers a different colatitude from previous experiments (BNL, Fermilab, CERN, J-PARC; see Table~\ref{tab:colatitude}). 
This distinct location not only enables complementary sensitivity in $\mu^+$/$\mu^-$ comparisons at the same site, but also allows cross-site combinations (e.g., FNAL $\mu^+$ with HIAF $\mu^-$) to probe additional linear combinations of SME coefficients that are otherwise inaccessible at a single latitude (see~\cite{Bluhm:1999dx} for details). 

\begin{figure}[ht]
\centering
\includegraphics[width=0.99\linewidth]{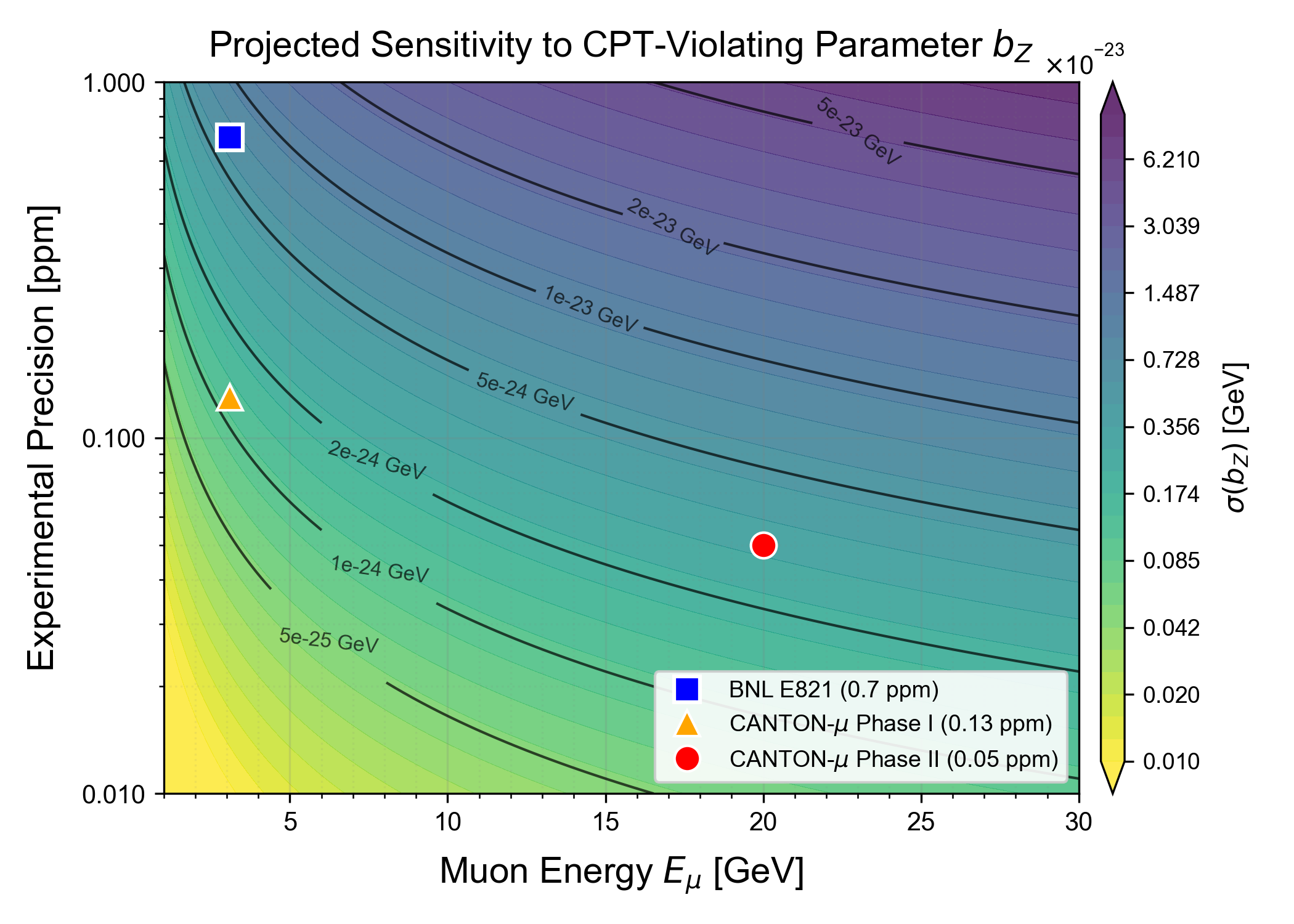}
\caption{\label{fig:NP_CPT}Projected sensitivity to CPT violation in the muon sector via the SME $b_Z$ parameter, as a function of muon energy and experimental precision. 
}
\end{figure}

Figure~\ref{fig:NP_CPT} illustrates the projected sensitivity to CPT-violating effects under the $b_Z$ parameter. 
For Phase~I, a 3~GeV muon beam with an experimental precision of $\sim 0.1$~ppm could probe $b_Z$ at the $\mathcal{O}(10^{-24})~\mathrm{GeV}$ level, representing an order-of-magnitude improvement over current limits. 
In Phase~II, higher muon energies reduce the sensitivity due to the $1/\gamma$ scaling, yielding slightly weaker constraints. Nonetheless, these measurements still provide stringent and complementary coverage of the SME parameter space and are therefore essential for next-generation CPT tests in the muon sector, with a sensitivity comparable to the Planck scale, given $m_\mu/M_{\rm P} \sim 10^{-21}$.

\begin{table}[ht]
\centering
\caption{Geographic latitude and colatitude of existing and proposed $g\!-\!2$ facilities.}
\begin{tabular}{lcc}
\hline
Facility & Latitude $\phi$ & Colatitude $\chi=90^\circ-\phi$ \\
\hline
BNL       & $40.9^\circ$N & $49.1^\circ$ \\
Fermilab  & $41.9^\circ$N & $48.1^\circ$ \\
CERN      & $46.2^\circ$N & $43.8^\circ$ \\
J-PARC    & $36.5^\circ$N & $53.5^\circ$ \\
HIAF      & $23.1^\circ$N & $66.9^\circ$ \\
\hline
\end{tabular}
\label{tab:colatitude}
\end{table}

\paragraph{Other potentials.} The flexible beam configurations enabled by the proposed facility, particularly the \rev{potentially} fast data-taking in Phase~II, also open windows to other new physics scenarios. A compelling example is the search for axion-mediated monopole-dipole forces~\cite{Agrawal:2022wjm} on muons, which would manifest as a spin precession contribution that is identical for $\mu^+$ and $\mu^-$ but changes sign when the beam direction is reversed from clockwise (CW) to counter-clockwise (CCW). This distinctive signature is accessible only through CW/CCW comparison and offers a clean probe of ultra-light axions with masses $m_\phi \lesssim 10^{-10}$ eV.
In particular, the axion interpretation of the current $g-2$ anomaly prefers a mass around $m_\phi \sim 10^{-12}$ eV (see Fig.~1 of~\cite{Agrawal:2022wjm}), a region where the improved precision of our proposed measurement (0.05~ppm in Phase~II) would enhance the sensitivity by approximately an order of magnitude.

Finally, we remark that the experimental concepts outlined here could also be adapted for a muon electric dipole moment (EDM) search, offering a complementary probe of the violation of the combined symmetry of charge and parity~\cite{Raidal:2008jk,Crivellin:2018qmi}. Whether the ultimate sensitivity of such a parasitic EDM program could become competitive to the current proposed muEDM experiment at PSI~\cite{Adelmann:2021udj} must be studied in detail.

\section{Conclusion and Outlook}
\label{sec:conclusion}

The CANTON-$\mu$ project explores the feasibility of a next-generation measurement of the muon anomalous magnetic moment at HIAF. \rev{Based on the projected beam-line performance and simulations of the resulting muon-beam properties, this study assesses the prospects for high-precision measurements with both $\mu^-$ and $\mu^+$.} Two preliminary storage-ring concepts are considered: a sector-magnet ring employing edge focusing and a full-ring configuration employing hybrid weak focusing. \rev{Each offers a possible route beyond some of the limitations of the conventional magic-momentum approach, while introducing distinct sources of systematic uncertainty and technical challenges that help define the priorities for future R\&D.}

\rev{The beam-line simulations provide the basis for the projected statistical sensitivities. The systematic-uncertainty estimates, by contrast, should be regarded as indicative design targets based on an assessment of the dominant uncertainty sources and their possible mitigation. Taken together, these projections indicate that CANTON-$\mu$ could achieve an overall precision of approximately $0.10~\mathrm{ppm}$ in Phase~I, comparable to that achieved by Fermilab E989. The proposed HIAF-U upgrade is expected to provide higher-energy muon beams with more than an order-of-magnitude increase in intensity. The enhanced beam conditions could enable Phase~II to reach an overall precision of approximately $0.05~\mathrm{ppm}$.} This would improve upon the present Fermilab benchmark by a factor of approximately 3 and provide an independent measurement with complementary experimental systematics.

Such precision would also offer significant physics opportunities.
Together with anticipated improvements in the Standard Model prediction,
a measurement at the $50~\mathrm{ppb}$ level would extend the indirect
sensitivity to new physics, particularly in chirally enhanced scenarios
for which the mass scales probed may exceed the direct reach of current
collider searches. Moreover, high-precision measurements with both
$\mu^+$ and $\mu^-$ would allow for a sensitive test of CPT symmetry in the muon sector at the $\mathcal{O}(10^{-24})~\mathrm{GeV}$ level, improving existing bounds by an order of magnitude. Together, these features define a unique and largely unexplored physics reach.

A new muon $g\!-\!2$ experiment will require a decade-scale
development of accelerator, storage-ring, detector, magnetometry, and
beam-dynamics studies. The present study therefore does not
constitute a finalized experimental design; rather, it identifies the
available beam opportunity, develops two candidate measurement concepts,
and establishes quantitative performance goals against which future
R\&D can be assessed. The next steps include end-to-end beam and spin
tracking, realistic field and lattice modelling, prototype development
of the principal instrumentation, and experimental validation of the
proposed systematic-control strategies. These studies would provide the
technical basis for a future Conceptual Design Report and, ultimately,
a full experimental proposal for a new generation of precision muon
$g\!-\!2$ measurements.

\section*{Acknowledgments}

The authors thank James Mott, David Hertzog, Kim Siang Khaw,
Paolo Girotti, Elia Bottalico, Andrzej Wolski and Martin Fertl for valuable
discussions on the experimental aspects of this work. The authors are also grateful to Graziano Venanzoni, Yannis Semertzidis, Alberto Lusiani, Thomas Teubner, and Dominik Stöckinger for their helpful comments, advice, and encouragement.

C.~Z. and F.~I. are supported by the Leverhulme Trust under Grant LIP-2021-014. O.~K. is supported by the U.S. DOE under Contract No. DE-SC0011665. Q.~L. is supported by the National Natural Science Foundation of China under Grant No. 12325504. B.~L. is supported by the National Natural Science Foundation of China under Grant No. 12305217. Z.~L. and L.~L. are supported by the National Natural
Science Foundation of China under Grant No. 1247052577. This work is also supported by the State Key Laboratory of Nuclear Physics and Technology at Peking University (Grant No. NPT2025KFY07) and the
Science and Technology Facilities Council (STFC), UK.

%%%%%%%%%%%%%%%%%%%%%%%%%%%%

%
% BibTeX users please use
% \bibliographystyle{}
% \bibliography{}
%
\bibliographystyle{unsrt}
\bibliography{canton2}
%\begin{thebibliography}{}
%Author, \textit{Book title} (Publisher, place year) page numbers
% etc
%\end{thebibliography}

\end{document}